\documentclass[aps,pre,twocolumn,superscriptaddress]{revtex4-1}

\usepackage{hyperref}
\usepackage{amsmath, mathrsfs, amssymb, wasysym}
\usepackage{graphicx}
\usepackage[caption=false]{subfig}

\begin{document}


\title{Memory Effects in Schematic Models  of Glasses Subjected to Oscillatory Deformation}


\author{Davide Fiocco}
\affiliation{Institute of Theoretical Physics (ITP), Ecole Polytechnique F\'ed\'erale de Lausanne (EPFL), 1015 Lausanne, Switzerland}
\author{Giuseppe Foffi}
\email{giuseppe.foffi@u-psud.fr}
\affiliation{Laboratoire de Physique de Solides, UMR 8502, B\^{a}t. 510, Universit\'e Paris-Sud, 91405 Orsay, France}
\author{Srikanth Sastry}
\email{sastry@jncasr.ac.in}
\affiliation{Jawaharlal Nehru Centre for Advanced Scientific Research, Jakkur Campus, Bangalore 560 064, India}
\affiliation{TIFR Centre for Interdisciplinary Sciences, 21 Brundavan Colony, Narsingi, 500075 Hyderabad, India}


\date{\today}


\begin{abstract}
We consider two schematic models of glasses subjected to oscillatory shear
deformation, motivated by the observations, in computer simulations of
a model glass, of a nonequilibrium transition from a localized to a
diffusive regime as the shear amplitude is increased, and of
persistent memory effects in the localized regime. The first of these
schematic models is the $NK$ model, a spin model with disordered multi-spin
interactions previously studied as a model for sheared amorphous
solids.  The second model, a transition matrix model, is an abstract
formulation of the manner in which occupancy of local energy minima
evolves under oscillatory deformation cycles. In both of these models,
we find a behavior similar to that of an atomic model glass studied earlier. 
We discuss possible further extensions of the approaches outlined. 
\end{abstract}

\maketitle

\section{INTRODUCTION}

Most mechanical systems are constantly subject to deformation during
their life time. Under certain conditions, such deformations can
profoundly alter the microscopic and macroscopic properties such as,
in metallurgy, the strength (work hardening) or the ductility (strain
softening). Understanding the effect of such deformations is clearly
of great interest from a practical point of view and, at the same time, poses many
fundamental questions~\cite{ashby2005materials}. Among different systems, amorphous materials are
particularly challenging from a conceptual point of view, due to the
lack of microscopic long-range order~\cite{alexander1998amorphous, Berthier2011, berthier2011theoretical,rodney2011modeling,falk2011deformation}. They are, however, very suitable for many practical applications  and arise in many context in nature.  A prominent example is the case of metallic glasses, which are disordered
solids that are obtained by fast cooling metallic alloys that have been especially designed to avoid crystallization and remain
amorphous~\cite{argon1979plastic,ashby2006metallic, chen2011brief}. Other widely studied examples, in addition to the most familiar class of molecular and polymeric glass formers, are colloidal suspensions~\cite{schall2007structural,eisenmann2010shear}, foams~\cite{VanHecke2010}, granular packings~\cite{Keim2014,Keim2013} and biological assemblies such as the cytoskeleton~\cite{Bursac2005}. 

Insight into the behavior of glass formers can be obtained by a computational investigation of their energy landscape~\cite{stillinger1995topographic,sastry2001relationship,Debenedetti2001, sciortino2005potential}. The same idea can be applied to mechanical deformation of glasses. In this case, the evolution of the local energy minima, or {\it inherent structures}, of the model glass, can be followed using a protocol referred to as {\it athermal quasi-static (AQS)} deformation \cite{maloney2006amorphous}.  It has been shown, for example, that systems tend to visit energy basins with energies typical of high temperature \cite{utz2000atomistic} under shear deformation up to large strains, while under a finite amplitude cycle of back and forth deformation, both rejuvenation (increase in the inherent structure energy) and over-aging (decrease in the energy) effects can be observed \cite{lacks2004energy}.

Recently, it has been shown that very rich and interesting behavior arises 
when a model glass is subjected to repeated cyclic deformations
at zero temperature, i.e. following the AQS protocol. For a model
binary mixture glass with particles interacting with the Lennard-Jones
potential~\cite{kob1995testing}, we have shown that as the amplitude of the oscillations
increases, the system undergoes a transition from a quiescent or
localized regime to a diffusive regime. In the former, after a short transient, the system
remains localized in the same energy minimum
at the end of each cycle, while in the latter state it diffuses in
configuration space~\cite{fiocco2013oscillatory}. This transition from localization to diffusion occurs at a critical
amplitude $\gamma_c$, as has also been reported by other authors on
similar systems~\cite{regev2013onset, priezjev2013heterogeneous},  in recent experiments~\cite{Knowlton2014,HimaNagamanasa2014} as well in other theoretical and computational works~\cite{Perchikov2014,Mosayebi2014}.  As discussed in~\cite{fiocco2013oscillatory}, the transition observed at $\gamma_c$
can be identified with the yielding transition under steady strain,
where irreversible behavior sets in. This transition resembles
strongly the dynamical transition from a reversible to an irreversible
state that has been found in dilute non-Brownian colloidal particle
suspensions \cite{corte2008random} and granular systems~\cite{Keim2014,Keim2014,Slotterback2012,Royer06012015}, but the
similarities and differences in the yielding behavior of these systems
merits further investigation.  In
that case, however, the reversible states, analogous to our localized
states, are a consequence of the intrinsic reversibility of the
low-Reynolds number hydrodynamics and they disappear above a certain
critical threshold when the interactions between the particles sets
in.
\\
The existence of reversible or localized states, that remain unchanged
under the effect of oscillatory deformation, implies that the system
remains indefinitely in a given state that
corresponds to the amplitude of the oscillatory deformation
imposed. In principle, this property can be used to encode memory that
can be read at a later stage. 
Keim an Nagel have used a computational model, first introduced in \cite{corte2008random}, to demonstrate that such a memory encoding is possible for a model of dilute colloidal particles~\cite{keim2011generic}.
 In the case of localized states, the system retains memory of the value of the amplitude that
has been used during the training phase and this value can be read by
analyzing the displacement during a deformation cycle, i.e. the
reading phase. We have shown that surprisingly analogous behavior can
be observed for a model atomic glass \cite{fiocco2014encoding}, though
with important differences. Glasses possess a complex energy
landscape and even when the microscopic configuration remains
invariant at the end of a full cycle of deformation, the system may
move through a complex periodic orbit of energy minima during the cycle. In addition, by alternating strain cycles of different amplitudes, multiple memories of such amplitudes can be encoded. Such a
possibility exists also in the model colloidal suspension but in
that case the ability to encode multiple memories is transient~\cite{keim2011generic}, and is observed
when the system is not in the fully trained, stationary, regime, and
requires the addition of noise to be made persistent for a large number of training cycles. 
Understanding the reasons why these (and other) significantly
different systems exhibit very similar memory effects, and the nature
of the universalities and differences is an interesting subject for
further investigations. In this paper, we describe results concerning two simple,  schematic models, which have been investigated to elucidate the characteristics and mechanisms of the non-equilibrium transition and memory effects described
above. These simplified models are investigated in order to understand what are the fundamental mechanisms that can originate the phenomenology that has been observed for the LJ system.
 The first is a lattice model, the {\it NK} model, that we have
already briefly discussed earlier in the context of memory
effects~\cite{fiocco2013oscillatory}. The NK model is a spin model
with multi-spin interactions that is designed to generate energy
landscapes with tunable roughness, and also to mimic the effects of
imposing shear deformation. The second model, the {\it transition
  matrix} (TM) model, aims to capture key aspects of the changes
induced by cyclic shear deformation by specifying rules to construct
a transition matrix that maps the set of local energy minima onto itself
as a result of a single cycle. Studying the evolution of the set of
occupied local energy minima as a function of the number of cycles
allows us, in principle, to understand what are the key features of the transition matrix that determines the 
observed behavior.

The paper is organized as follows. In \autoref{intro-NK} and
\autoref{intro-TM} we introduce the two models that are the main
subject of this paper. In \autoref{results-transition} we discuss the
results that concern the existence of a dynamical transition while in
\autoref{results-memory} we discuss memory effects. Finally, we
summarize our results and conclusions in \autoref{conclude}.

\section{THE NK MODEL}
\label{intro-NK}

The NK model is defined on a set of spins interacting according to a Hamiltonian
that incorporates a parameter $\gamma$ that mimics the effect of shear strain. 
It was studied by Isner and Lacks \cite{isner2006generic} as an analog
system to the model atomic glass former in which overaging and
rejuvenation were observed \cite{lacks2004energy} under a single
deformation semicycle.

We consider $N$ (even) lattice sites occupied by spins $m_{i}$ that can take either the values $0$ or $1$.
\begin{equation}
\{m_1, m_2, \ldots, m_i, \ldots, m_{N} \} \in \{0, 1\}^{N}
\end{equation}
Furthermore, in order to prevent the system getting trapped around low energy configurations with aligned spins, we limit the space of allowed configurations to those that satisfy the constraint
\begin{equation}
\sum_{i} m_{i} = \frac{N}{2}
\end{equation}
(this is equivalent to taking the set of states of constant
magnetization $0$ in the Ising model). There are $\binom{N}{N/2}$ such
configurations, and we define two NK configurations as adjacent if one
is turned into the other by swapping the values at two sites $i$ and
$j$ such that $m_{i} \neq m_{j}$. Since one may choose one of $N/2$
spins with $m = 1$ for swapping with any one of $N/2$ spins with $m =
0$, each configuration has $N^{2}/4$ configurations that are adjacent
to it.

We introduce:
\begin{itemize}
	\item An ordered list of $K$ ``neighbors'' for each $i$-th spin, specified by the map $J$: 
	\begin{equation}
		m_{i} \xrightarrow{J} \{m_{i}^{1}, \ldots, m_{i}^{K}\}
		\label{eq:NKNeighbors}	
	\end{equation} 
	The choice of the list of neighbors for a given spin is random.
	\item Two functions $a$ and $b$ connecting the set $\{0, 1\}^{K+1}$ (the set of all the $(K+1)$-tuples formed by ones and zeros) to the intervals $[-1,1]$ and $[0,1]$ respectively
	\begin{align}
		\{0, 1\}^{K+1} & \xrightarrow{a} [-1,1]  & \quad		\{0, 1\}^{K+1} & \xrightarrow{b} [0,1]
		\label{eq:NKCouplings}
	\end{align}
	The correspondence between a given $(K+1)$-tuple and the numerical value is chosen randomly with a uniform probability within the respective intervals.
\end{itemize}
 
The energy of the system depends on the values of the spins, $E = E(m_{1}, \ldots, m_{N})$ according to
\begin{equation}
	E = -\frac{1}{2} \sum_{i = 1}^{N} \left[ 1 + \sin(2 \pi (a_{i} + \gamma b_{i})) \right]
	\label{eq:NKEnergy}
\end{equation}
where $\gamma$ (which we will name, with an abuse of language, ``shear strain'') is a parameter that can be varied continuously.

The NK model is known to possess a discrete energy landscape whose
roughness (a measure of which is given by the number of local minima
in the landscape) is tuned by the value of the parameter $K$. The
landscape is smooth for $K = 0$ and the roughness is increased as $K$
is increased. To see this consider the case $\gamma = 0$ for
simplicity. If $K=0$, when performing the sum in \autoref{eq:NKEnergy}
one sums contributions that are simply $1 + \sin(2 \pi a_{i})$, where
$a_{i}$ can assume only two values in the interval $[-1,1]$, depending
solely on whether $m_{i} = 1 $ or $0$, as no spin has neighbors. It is
clear that the energy increases or decreases monotonically with the
``magnetization'', $\sum_{i} m_{i}$. As in our case such quantity is
fixed, all the allowed configurations have the same $E$. This implies
that by moving from a given configuration to any adjacent one the
energy can't change and the landscape is thus \emph{flat}. In the case
$K = N-1$ each spin is a neighbor of any other. To estimate the
roughness of the landscape, we compare the energies of two adjacent
structures, that differ by the swapping of two spins. Each term in
\autoref{eq:NKEnergy} is affected by such a swap, and so all the
$a_{i}$'s that contribute to the energy of the two configurations
differ. Since the $a_{i}$'s are random and uncorrelated, there won't
be correlation between the energies of two adjacent
configurations. Consequently, the structure of the overall landscape
will be rough.

While $K$ is able to tune the roughness of the landscape, the
parameter $\gamma$ is able to change continuously the values of the
energy of the configurations (and thus the overall landscape). This is
similar to what happens in a glass, in which a
macroscopic strain is externally imposed.

The broad features of the NK model are thus similar to the atomic
glass formers mentioned earlier, with the significant difference that
the configuration space of the NK model is discrete. We next describe
how procedures such as energy minimization and AQS are performed in
the NK model.  Starting from an initial configuration, energy can be
minimized by a steepest descent (SD) procedure. A SD in the NK energy
landscape consists of moving from a configuration to the adjacent one
with the lowest $E$, and iterating this procedure until when no move
to an adjacent configuration results in a decrease in $E$. Using such
a protocol, any configuration can be mapped onto a local minimum,
{\it i.e.} an inherent structure of the NK landscape. This fact, together
with the dependence on $\gamma$ of the landscape, allows us to define an
athermal quasi static ``deformation'' procedure on NK systems too:

\begin{enumerate}
	\item Take an inherent structure of the NK landscape. This can be obtained by any configuration by means of the SD algorithm.

	\item Increment the value of $\gamma$ in \autoref{eq:NKEnergy}
          by a small amount $d\gamma$. The value of $d\gamma$ should
          be small enough so that the modification of the landscape is
          slow enough and no displacements to adjacent inherent
          structures are ``missed''\footnote{In other words, $d\gamma$
            should be small enough that if a smaller $d\gamma'$ was
            employed it would yield exactly the same evolution.} by
          the AQS dynamics.

	\item Apply the SD procedure to the configuration.
\end{enumerate}

As AQS can be applied to the NK model, it is interesting to check
whether the same phenomenology seen in AQS deformation of model atomic
glass formers can be observed in it. However, one should bear in mind
the following differences between the NK model and atomic glass
formers: 

\begin{itemize}
	\item Since the NK model has a \emph{discrete} configuration
          space, the AQS dynamics of NK systems is somewhat different to
          that of atomic systems under the AQS protocol. In the NK
          case, the system occupies one given point of the available
          configuration space as $\gamma$ is changed, and stays there
          until it ``jumps'' to another inherent stucture
          configuration as soon as the initial configuration is not an
          inherent structure anymore. In the atomic systems, instead,
          the configuration of the system \emph{continuously changes} as
          $\gamma$ is varied.

	\item Due to the discrete nature of the landscape,
          minimization is trickier in the NK case. While for atomic
          glass formers \emph{local} quantities (e.g. the calculation
          of a potential energy gradient) allow the determination of
          directions to follow to reach an inherent structure, in the
          NK model \emph{all} the energies of adjacent configurations
          need to be calculated in order to choose the adjacent
          configuration with the lowest energy (if there is one). This
          operation requires $O(N^{2})$ energy calculations to be
          performed (as the number of adjacent configurations scales
          with $N^{2}$, see above) and it is thus computationally
          infeasible for large values of $N$.

	\item While for atomic glass formers, the energy landscapes
          depend on the few parameters defining the interaction
          potential ({\it e. g.} the values of the $\epsilon$ and
          $\sigma$ parameters for the Lennard-Jones potential)
          and the boundary
          conditions of the simulation volume, the definition of the
          landscape in the NK case requires the introduction of a much
          larger number of parameters. This is because the lists of
          neighbors specified by $J$ can be realized in many different
          ways and $a$ and $b$ in
          \autoref{eq:NKNeighbors} and \autoref{eq:NKCouplings}
          require $2^{K}$ values each to be defined.
\end{itemize}

Having noted this, we can study the NK model under athermal
quasi-static deformation, and compare the results with what is
obtained with the model atomic glass former studied in
\cite{fiocco2013oscillatory} (the Kob-Andersen binary mixture with
Lennard-Jones interactions~\cite{kob1995testing}, or LJ). Such comparison is meaningful,
because the two systems share important features. At the same time it
is not trivial, becuase, as explained above, the two models are
``sufficiently different'' that common qualitative behavior can not be
taken for granted {\it a priori}.

\begin{figure}[!h] 
\centering 
\includegraphics[width=0.5\textwidth]{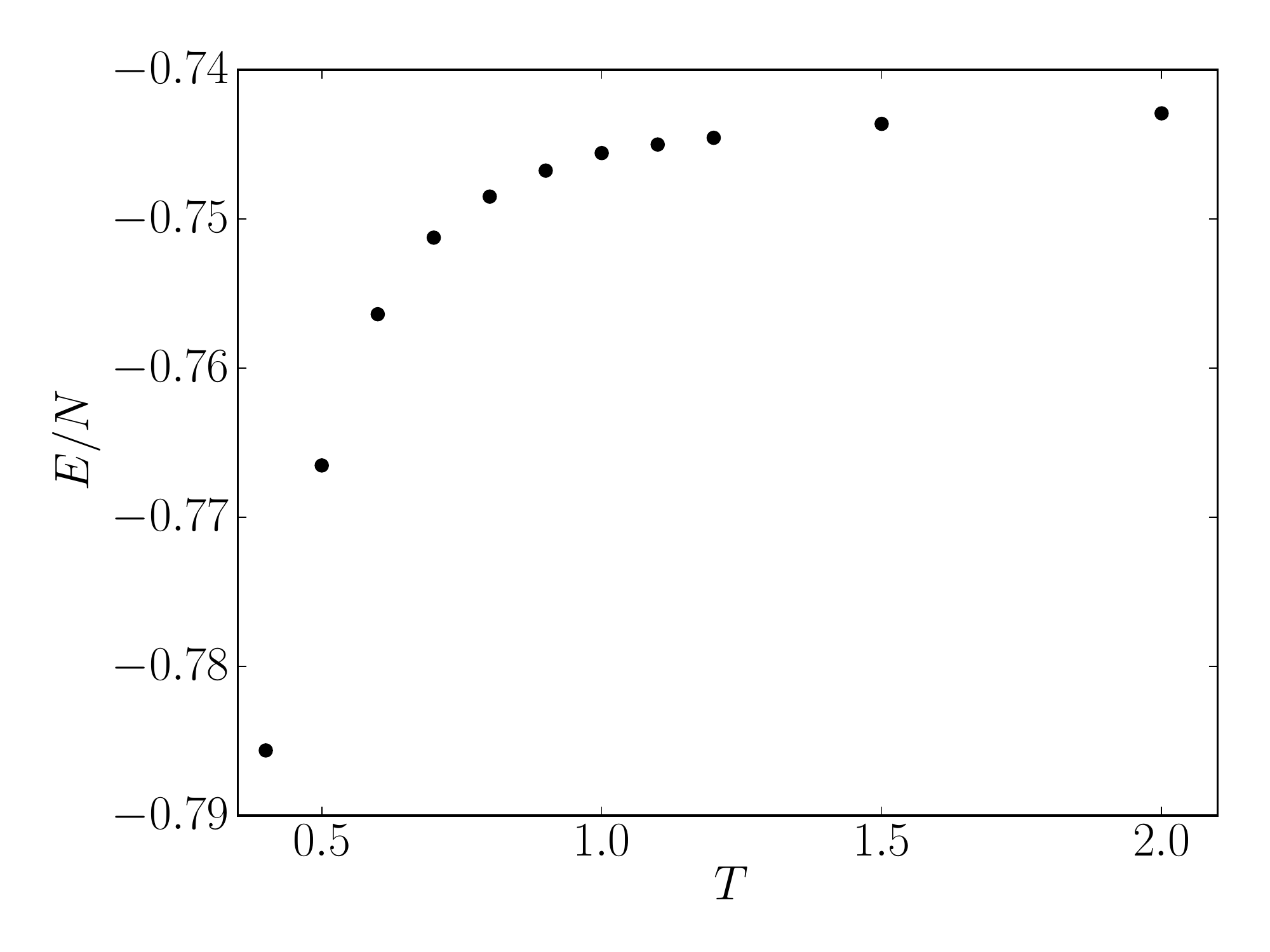} 
\caption{Average energy per site of the inherent structures obtained by quenching NK samples ($N = 20$, $K=10$) equilibrated at different $T$.\label{fig:InherentUvsTNK}}
\end{figure}

Before performing simulations of ``deformation'' on NK samples one
needs initial configurations.  As in the case of the LJ
systems discussed in \cite{fiocco2013oscillatory}, one would like to
start from sets of configurations that differ for some feature (like
the average potential energy). In this way one can check how
deformation affects the samples, and how it is capable of making them
evolve in such a way to forget their initial state. The obvious choice
is to choose inherent structures corresponding to different
temperatures $T$, as we did in our previous study of the LJ case. We thus obtain
several equilibrated NK configurations at different $T$. We do this
for several realizations of the couplings $a$, $b$ and $J$ by means of
Monte Carlo sampling and find the inherent structures at each
temperature $T$ by means of steepest descent (SD) minimization. The
average energy of such inherent structures is plotted in
\autoref{fig:InherentUvsTNK}, showing a behavior similar to that
observed in \cite{sastry1998signatures} for a model atomic glass former. We choose temperatures (measured in units of $k_B=1$) $T = 0.6$
and $1.0$ for further analysis, as they are different enough to be
distinguished. These temperatures are above the glass temperature of the models which is estimated to be around 0.45~\cite{isner2006generic}.

\section{Transition matrix model \label{sec:TransitionMatrixModel}}
\label{intro-TM}

Both the LJ and NK models possess a rugged landscape that is
modified during the deformation. The behavior under oscillatory
deformation, in particular, thus depends on the detailed features of
the landscape.  Would it be possible to predict the behavior of such
models qualitatively, without encoding in detail the features of the
energy landscape, but using a sort of ``high-level'', abstract
description of its evolution? This is what the ``transition matrix''
method (TM) aims to do. 

The starting point of the TM approach is to look at a cycle of AQS
deformation from a mere ``mathematical'' point of view.  In that
perspective, an AQS cycle is a correspondence between the set of $M$
inherent structures of the energy landscape at $\gamma = 0$ into
itself. This is because a valid starting configuration is an inherent
structure of the $\gamma = 0$ landscape, and it is transformed into
another inherent structure of the same landscape at the end of the
deformation cycle. Each of these inherent structures can be identified
by an index $i$, and associated to a $M$-dimensional vector
$\mathbf{R_{i}}$ whose components are all zero but for the $i$-th one,
which is set equal to 1. Then, they can be taken as starting points of
a deformation experiment where a single AQS cycle is performed and the
inherent structure reached at the end is recorded.

This allows to define a \emph{transition matrix} $P$, such that
\begin{equation}
	P \mathbf{R_{i}} = \mathbf{R_{f}}
	\label{eq:TransitionMatrixCycle}
\end{equation}
where $\mathbf{R_{i}}$ and $\mathbf{R_{f}}$ are the vectors associated
to the initial and final inherent states.  Here we list some of the
properties of the $P$ matrix, which derive directly from the features
of AQS dynamics:
\begin{itemize}
	\item $P$ encodes the entire information about the evolution of any inherent structure under cyclic deformation, as by using \autoref{eq:TransitionMatrixCycle} one can determine the final inherent structure $\mathbf{R_{f}}$ given any intial one $\mathbf{R_{i}}$.
	\item $P$ is a sparse $M \times M$ matrix, and $P_{ij} = 1$ if and only if the state associated to $\mathbf{R_{j}}$ is mapped onto the inherent structure $\mathbf{R_{i}}$ in the AQS cycle. The consequence is that all the columns have exactly one non-zero entry (equal to one) because each $\mathbf{R_{j}}$ configuration is sent to some other $\mathbf{R_{i}}$ inherent structure by the AQS cycle.
	\item $P$ depends on the value of $\gamma_{max}$, i.e. on the amplitude of the deformation. For very small amplitudes, a sizable fraction of the inherent structures will be unchanged under the deformation, because an AQS cycle will not be effective at destabilizing the starting inherent structures. In this case a given structure will often map onto itself through $P$ and thus $P$ will be very close to the diagonal unit matrix. In general this won't be any longer true for higher values of $\gamma_{max}$.
	\item The determinant of $P$ is, in general, zero. In general, more than one structure will map onto the same final state $\mathbf{R_{f}}$, so that in that case $P$ will define a non-injective function. As the domain of $P$ is also its codomain, $P$ is not surjective, so that there will be structures that are not arrival configurations for any inherent structure (this is illustrated in \autoref{fig:AQSCycleAsMap}). This has the consequence that some rows of the $P$ matrix, in general, are identically zero, and so is the value of $\det P$.
	
	\begin{figure}
	\centering 
	\includegraphics[width=0.5\textwidth]{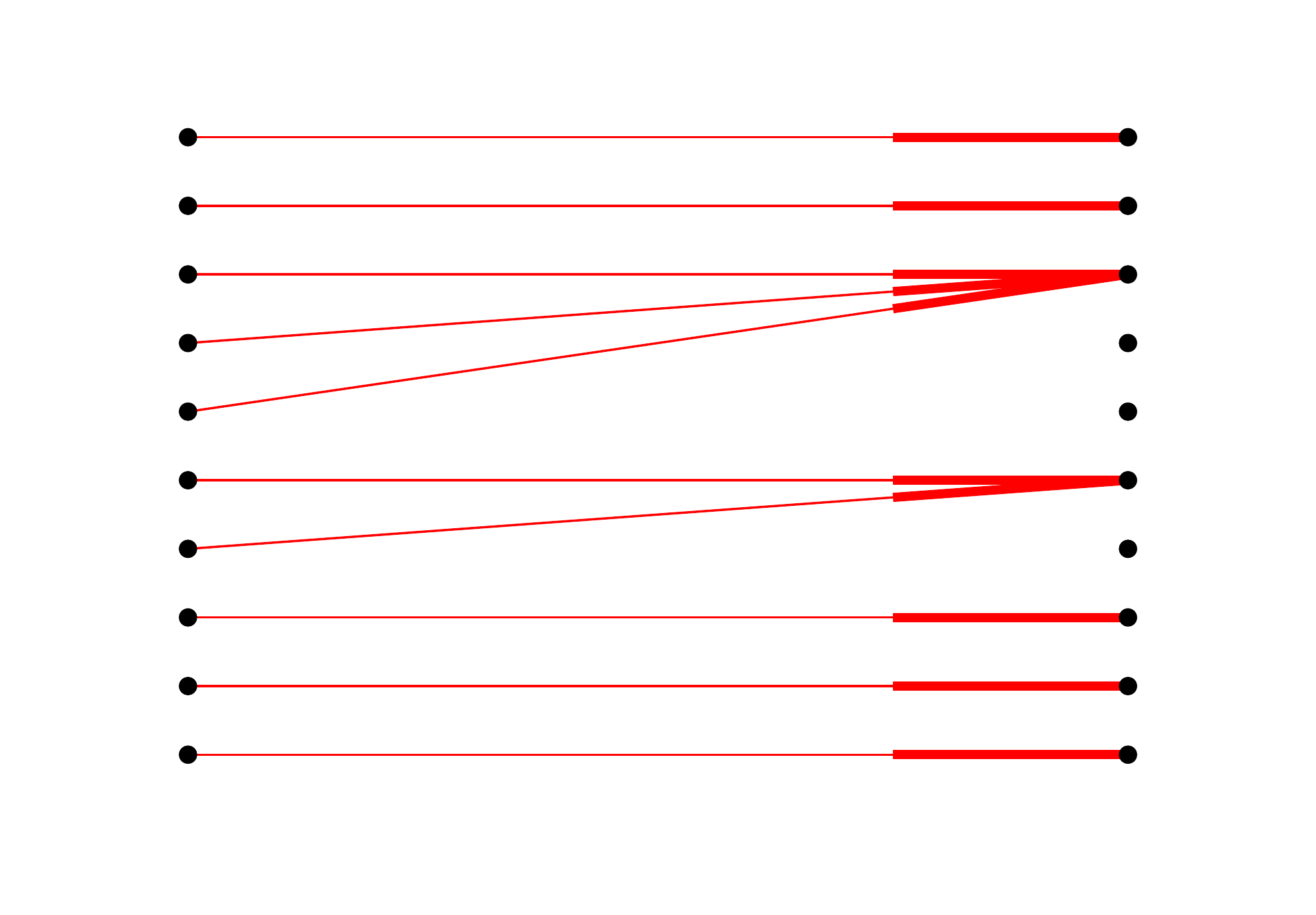} 
	\caption{Graph representation of an AQS cycle, that has the effect of mapping the set of inherent structures of the $\gamma = 0$ landscape onto itself.\label{fig:AQSCycleAsMap}}
	\end{figure}
	
	\item $M$, in general, is a large number. The size of $P$ is equal to the number of inherent structures of the landscape, and this number, for typical LJ and NK landscapes, is large (exponential in $N$ in the first case, and equal to $\binom{N}{N/2}$ in the second).
	\item The result of a deformation experiment where $L$ cycles (rather than just one) are applied to a given starting configuration $\mathbf{R_{i}}$ is obtained by applying \autoref{eq:TransitionMatrixCycle} repeatedly:
	\begin{equation}
		P^{L} \mathbf{R_{i}} = \mathbf{R_{f}} \qquad \text{where } P^{L} = \underbrace{P \cdot \ldots \cdot P}_{L\text{ times}}
	\end{equation} 
\end{itemize}

\subsection{Classification of states by their transformation properties \label{sec:TMStateClassification}}

A given configuration $\mathbf{R_{i}}$ can transform under the effect of $P$ in different ways:
\begin{enumerate}
	\item \label{it:AbsorbingState} $P \mathbf{R_{i}} = \mathbf{R_{i}}$: in this case $P$ has no effect on $\mathbf{R_{i}}$, so that $\mathbf{R_{i}}$ is an eigenvector of $P$ relative to the eigenvalue 1, and $P_{ii}=1$. We name such a $\mathbf{R_{i}}$ an \emph{absorbing} state.
	\item \label{it:RecurringState} $P^{L} \mathbf{R_{i}} = \mathbf{R_{i}}$ for some $L > 1$: in this case the oscillatory deformation starting from  $\mathbf{R_{i}}$ makes it cycle through a sequence of states, and after $L$ cycles $\mathbf{R_{i}}$ is reached again. $\mathbf{R_{i}}$ an eigenvector of $P^{L}$ relative to the eigenvalue 1. We name such a $\mathbf{R_{i}}$ a \emph{recurring} state.
	\item \label{it:MappingAnAbsorbingState} $P^{J} \mathbf{R_{i}} = \mathbf{R_{\rm abs}}$ for some $J \geq 1$, where $\mathbf{R_{\rm abs}}$ is an absorbing state. We name such a $R_{i}$ as \emph{mapping to absorbing state}.
	\item \label{it:MappingARecurringState} $P^{J} \mathbf{R_{i}} = \mathbf{R_{\rm rec}}$ for some $J \geq 1$, where $\mathbf{R_{\rm rec}}$ is a recurring state. We name such a $R_{i}$ as \emph{mapping to recurring state}.
\end{enumerate}

Note that \emph{every} configuration falls in one of the categories
enumerated above.  This can be demostrated as follows: suppose that a
configuration $\mathbf{X}$ exists that doesn't fall in any of the
categories listed above. If $P$ is applied to it, then some other
configuration $\mathbf{Y_{1}} \neq \mathbf{X}$ is obtained, else
$\mathbf{X}$ would belong to the category in
\autoref{it:AbsorbingState}. Repeated application of $P$ yields always
new states $\mathbf{Y_{2}}, \mathbf{Y_{3}}, \ldots$, so that no
absorbing nor recurring states are encountered, else $\mathbf{X}$
would belong to one of the categories in \autoref{it:RecurringState},
\autoref{it:MappingAnAbsorbingState} or
\autoref{it:MappingARecurringState}. After $M$ applications of $P$, a
sequence of $M+1$ distinct inherent configurations has been generated,
but only $M$ distinct inherent structures exist! So the initial
hypothesis (the very existence of $\mathbf{X}$) can't be true. We note
that in principle, recurring states with periods that are of
\cal{O}(1) are qualitatively different from and should be distinguished from those with are of \cal{O}(M),
although we do not attempt such analysis here.

What is the correspondence between this classification of states and
the absorbing and the diffusive states encountered in
\cite{fiocco2013oscillatory}? Absorbing states of the LJ model clearly
correspond to the absorbing states of the TM model. Diffusing states
in the LJ model correspond to a subset of the recurring states of the
TM picture with periods which are large, or more precisely, exponential in the number of particles in the system. In fact, even though a
diffusing LJ system does not seem to revisit the same inherent
structure as it travels in configuration space, after a large number
of oscillation cycles it \emph{has to}, as the number of possible
inherent structures is finite. Thus the diffusing states of can be
viewed as recurring states, which just take a \emph{very} large number
of cycles to come back to their starting state. Nevertheless, a more sophisticated analysis than we attempt here should distinguish between diffusing states and recurring states with \cal{O}(1) period cycles. 

\subsection{Construction of the $P$ matrix}
\autoref{eq:TransitionMatrixCycle} tells that the $P$ matrix contains the entire information about the outcome of
oscillatory AQS deformation, but how can one construct it?  Computing
it for LJ or NK systems can in principle be done by brute force: one
needs to have a list of the inherent structures, use each of them as a
starting configuration for a shear deformation cycle and see which
structures they eventually reach at the end of the cycle. This idea is
easier to apply for the NK model than in the LJ case: in the former
one can (at least in principle) enumerate all the $\binom{N}{N/2}$
allowed configurations, minimize each and every one of them so to get
all the inherent structures of the landscape; in the latter, which has
a continuous energy landscape and a continuous set of configurations,
the determination of the local minima of the landscape is a less
trivial task\footnote{Also in the NK case, however, this brute force
  approach to the determination of $P$ is very expensive
  computationally, as $\binom{N}{N/2}$ minimizations need to be
  performed at every AQS step.}.  The infeasibility of a brute force
approach is the reason why one would like to construct $P$ by less
expensive means, albeit in an approximate or schematic way. To do so, we make a
series of observations and assumptions about the evolution of the
energy landscape, that allow to construct $P$.

\subsubsection{Assumptions for constructing transition matrices \label{sec:TMAssumptions}}

The transition matrices that we construct are based on some assumptions on
the dynamics of inherent structures in the course of a deformation
cycle:

\begin{enumerate}
		\item During the deformation (at non-zero values of
                  $\gamma$) the number of energy minima is assumed to
                  be \emph{always} equal to $M$, no matter the value
                  of $\gamma$. The number of local minima present in
                  the landscapes of LJ and NK models will, in general,
                  weakly\footnote{A weak dependence of the number of
                    inherent structures in a Lennard-Jones system can
                    be justified by the fact that one does not expect
                    that the physics of a system depends on the
                    boundary conditions. The detailed energy landscape
                    of the system will depend on the value of $\gamma$
                    in a simulation that employs Lees-Edwards
                    conditions, but the \emph{statistical} properties
                    of the system (like the value of $M$) will not
                    vary much with it.} depend on the value of
                  $\gamma$.
		\item Minima are destabilized by changing $\gamma$,
                  {\it i.e.} that some of them are destroyed by the
                  deformation. As the number of structures is assumed
                  to be conserved (see above) to each inherent
                  structure destruction corresponds the creation of a
                  new one.
		\item The probability per unit strain of an inherent
                  structure to be destabilized is assumed to be the
                  same for all the $M$ structures and equal to a value
                  $\tau$ independent of $\gamma$.
		\item A system resides in a given inherent structure
                  until such inherent structure is destabilized. When
                  this happens, the system jumps to another inherent
                  structure of the deformed landscape. For simplicity
                  such a structure is assumed to be picked at random
                  in the landscape (whereas, in a realization of the
                  NK or LJ models, a system will land on a structure
                  which is not far from the starting one in the space
                  of configurations).
\end{enumerate}

In addition, the model relies on two facts that are true in general:

\begin{enumerate}
		\setcounter{enumi}{4}
		\item As a deformation \emph{semi}cycle brings the
                  system from 0 up to $\gamma_{max}$ and then back to
                  the undeformed landscape, there is a \emph{symmetry}
                  in the structures that are created and destroyed as
                  $\gamma$ is incremented from $0$ to $\gamma_{max}$
                  and those that are created/destroyed as $\gamma$ is
                  reduced back to 0 in the second part of the
                  semicycle. In fact, if a structure $\mathbf{R}$ is
                  destroyed when incrementing $\gamma$ above some
                  value $\gamma^{*}$, the same structure $\mathbf{R}$
                  will be created at $\gamma^{*}$ as the deformation
                  is reversed. The converse is true for a structure
                  $\mathbf{S}$ that is created in the first half of
                  the semicycle.
		\item The matrix $P$ can be viewed as the product of
                  $P_{+}$ and $P_{-}$, the matrices that describe the
                  two semicycles (one denoted by positive, the other
                  by negative strain $\gamma$) that form a full
                  oscillation cycle.
\end{enumerate}

A description of how these assumptions and observations are combined
to calculate an approximation of the $P$ matrix is presented in
Appendix \ref{append}.


\begin{figure}
\centering 
\includegraphics[width=0.5\textwidth]{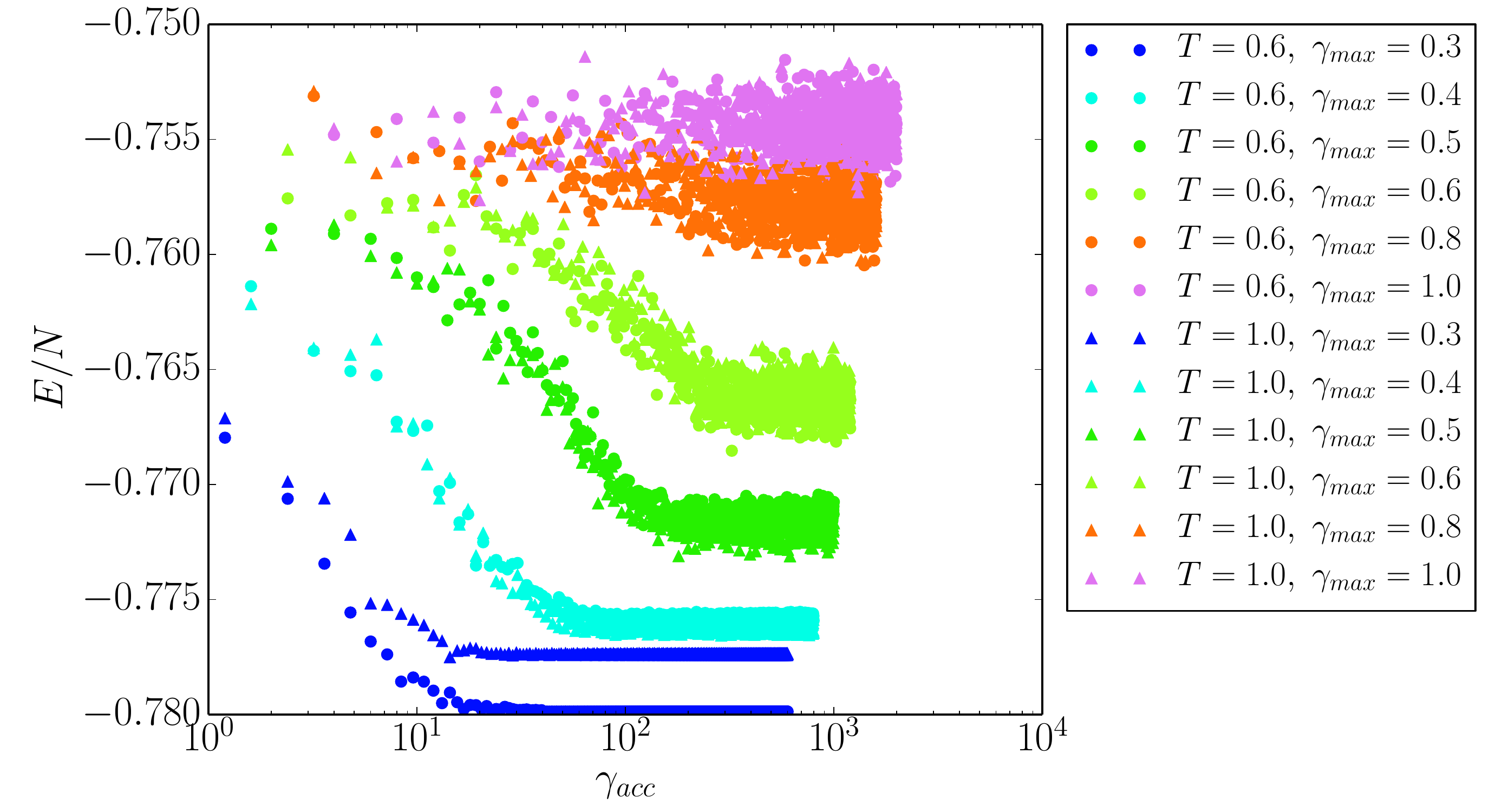} 
\caption{Potential energy per site as a function of $\gamma_{acc}$, for different initial effective temperatures and different deformation amplitudes $\gamma_{max}$, setting $N=40$, $K=10$. Data refer to configurations with $\gamma = 0$. For large values of $\gamma_{max}$ the energy fluctuates around some value which depends on $\gamma_{max}$ only. At small $\gamma_{max}$, instead, the plateau value of the energy depends on the effective $T$ of the initial configuration. In this respect what is observed here qualitatively resembles what has been found in the atomistic glass model in \cite{fiocco2013oscillatory}. \label{fig:UvsAccumulatedStrainNK}}
\end{figure}

\section{RESULTS: dynamical transition}
\label{results-transition}

\subsection{Dynamical transition under oscillatory deformation:  the NK model}

\begin{figure}
\centering 
\includegraphics[width=0.5\textwidth]{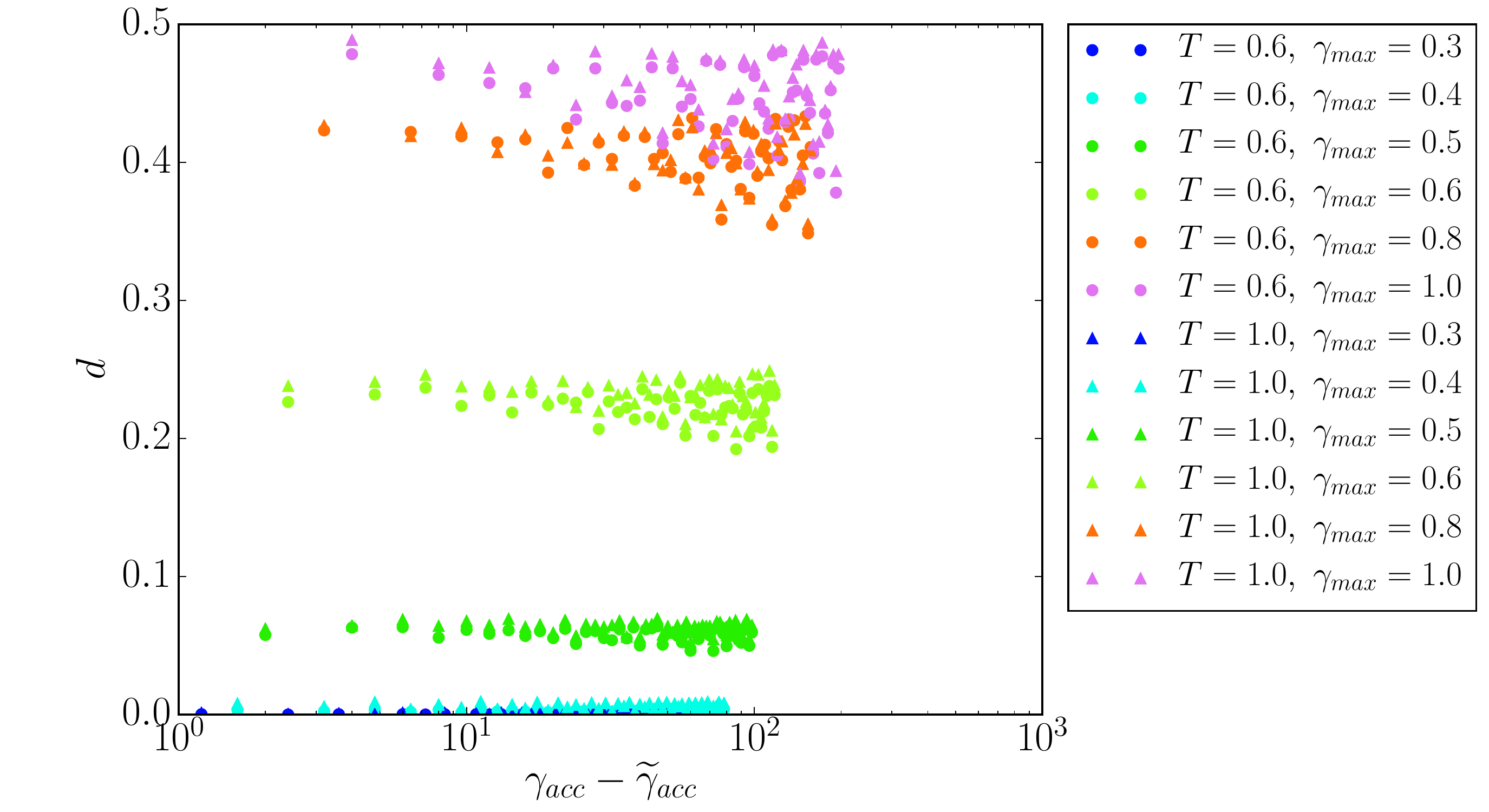} 
\includegraphics[width=0.5\textwidth]{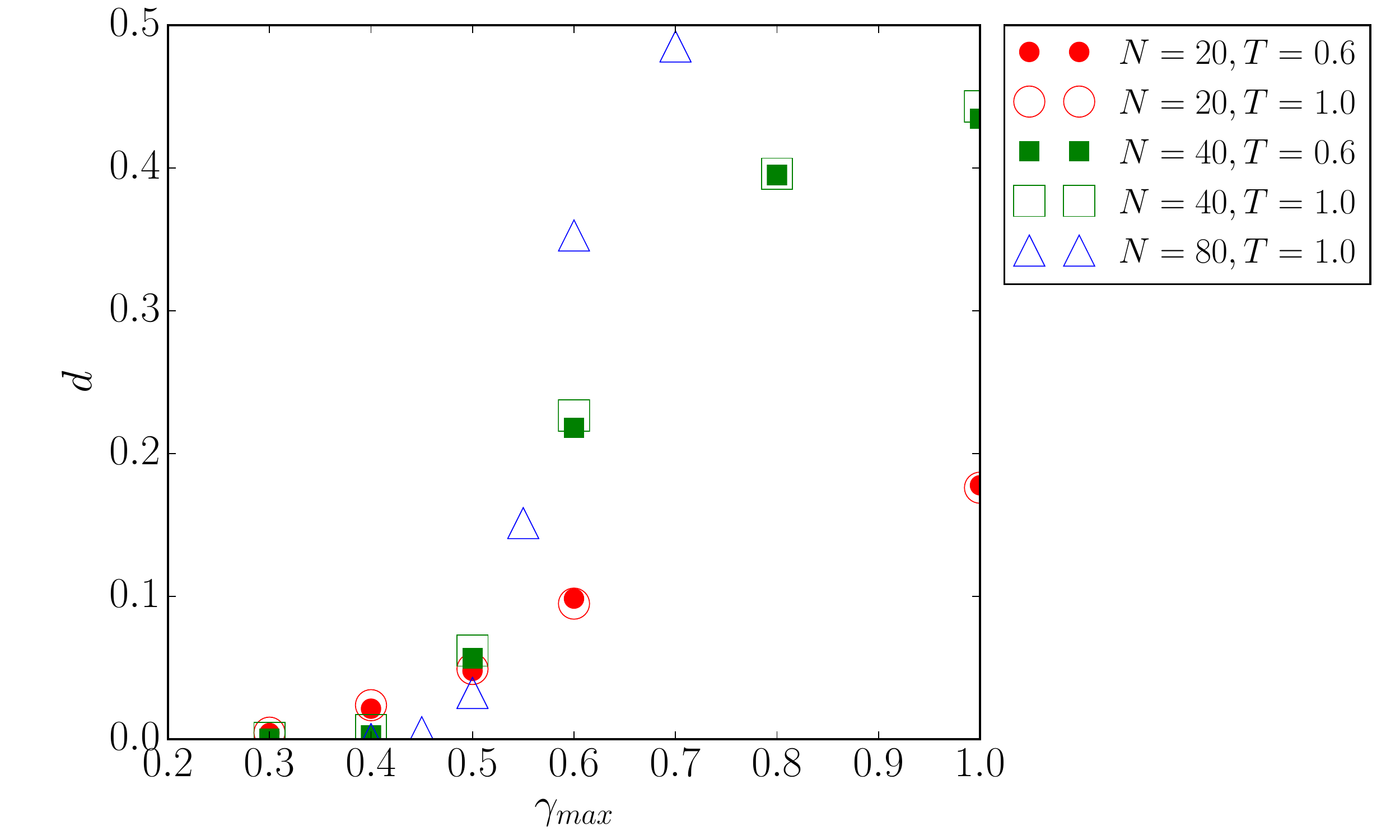}
\caption{(top) Hamming distance as a function of the accumulated strain measured from reference configurations with $\gamma_{acc} > \widetilde{\gamma}_{acc}$, where $\widetilde{\gamma}_{acc}$ marks the reaching of the plateau of the energy in \autoref{fig:UvsAccumulatedStrainNK}. $N=40$ and $K=10$. The behavior is not diffusive, but the higher the $\gamma_{max}$, the further the systems are able to move away from the reference configuration. The Hamming distance can be modeled with a constant function of $\gamma_{acc} - \widetilde{\gamma}_{acc}$\label{fig:HammingvsAccumulatedStrain}. (bottom) Average value of the Hamming distance as determined by averaging data like that in \autoref{fig:HammingvsAccumulatedStrain}, for different $N$ and setting $K=10$. The ability of the system to diffuse away from a reference configuration increases strongly with $\gamma_{max}$ at $\gamma_{max} \approx 0.5$. The transition becomes sharper as $N$ is increased. \label{fig:AverageHammingvsGammaMax}}
\end{figure}

We consider $N = 20, 40, 80$ and $K=10$, and $\approx 200$ instances of the couplings and obtain 3-4 equilibrated configurations at $T = 0.6$ and $1.0$ for each of such instances. The corresponding inherent structures obtained by SD are deformed by increasing $\gamma$ in \autoref{eq:NKEnergy} in steps of $d\gamma = 0.005$ (for all $N$) and performing SD at each step. The parameter $\gamma$ is varied in the interval $[-\gamma_{max}, \gamma_{max}]$ in a triangle wave fashion, exactly as in the AQS simulations seen in Ref.~\onlinecite{fiocco2013oscillatory, fiocco2014encoding}. The values of $E$ and the configuration are recorded whenever $\gamma = 0$, i.e. at intervals of $2\gamma_{max}$. Plots of $E$ as a function of the accumulated strain $\gamma_{acc}$  are shown in \autoref{fig:UvsAccumulatedStrainNK}  for the $N = 40$ case. Similarly to what is observed in the LJ case, for small values of $\gamma_{max}$ the energy reaches a plateau which depends on \emph{both} $\gamma_{max}$ and the initial effective $T$. For higher values of $\gamma_{max}$, all samples, regardless their initial effective $T$, reach plateau depending on $\gamma_{max}$ only. In this respect, the NK model is able to reproduce qualitatively the same behavior found in LJ systems~\cite{fiocco2013oscillatory}, where samples forget about their initial preparation if the oscillation amplitude exceeds some value $\gamma_{c}$. 
\\ \\
Changes in configurations under deformation can be studied for NK configuration as well, by looking at the distance between configurations before and after the application of deformation cycles.
Consider two configurations, $\mathbf{R}(\widetilde{\gamma}_{acc})$ and $\mathbf{R}(\gamma_{acc})$, obtained for values of the accumulated strain equal to $\widetilde{\gamma}_{acc}$ and $\gamma_{acc}$ respectively, with $\widetilde{\gamma}_{acc} < \gamma_{acc}$. Their distance can be expressed using the Hamming definition\footnote{In simpler terms, $d$ is the fraction of disagreeing components of the two vectors $\mathbf{r_{1}}$ and $\mathbf{r_{2}}$.} $d$:
\begin{equation}
	d(\gamma_{acc} - \widetilde{\gamma}_{acc}) = \frac{c_{01} + c_{10}}{N},
	\label{eq:HammingDistance}
\end{equation}
where $c_{01}$ ($c_{10}$) is the number of occurrences such that the $i$-th component of $\mathbf{R}(\widetilde{\gamma}_{acc})$ and the $i$-th component of $\mathbf{R}(\gamma_{acc})$ are respectively equal to 0 and 1 (to 1 and 0). 
We thus pick a configuration $\mathbf{r_{1}}$, choosing a large enough $\widetilde{\gamma}_{acc}$ so that the corresponding $E$ in \autoref{fig:UvsAccumulatedStrainNK} has relaxed to a steady state. We then compute the Hamming distance from it for configurations reached for increasing values of $\gamma_{acc}$, and plot it in \autoref{fig:HammingvsAccumulatedStrain}.

The average Hamming distance measured starting from a reference
configuration in the steady state quickly reaches a constant value for
increasing $\gamma_{acc}$, rather than show a linear increase with
time as the mean squared displacement does for the atomic glass former
\cite{fiocco2013oscillatory}. However, the Hamming distance, like the
overlap~\cite{Donati2002} or dissimilarity~\cite{Oppelstrup2005} used in the context of atomic liquids, is a bounded
quantity and should be treated as a correlation function rather than a
measure of mobility. In the present case, given the constraint on the total magnetization, the Hamming distance has an average value of $0.5$ for uncorrelated configurations. Thus, the value of the Hamming distance itself serves as a metric of localization of the configurations under shear. 

By plotting the average Hamming distance as a function of $\gamma_{max}$ for different values of $N$ (see \autoref{fig:AverageHammingvsGammaMax}), one observes that the capability of the system to move away from the reference configuration increases sharply at some value $\gamma_{c}$, which is roughly $N$ independent. The sharpness of the transition, moreover, is seen to increase with $N$. These data thus seem to confirm that a transition at some oscillation amplitude $\gamma_{c}$  from a localized regime to a diffusive one exists in the NK model, similarly to what is observed in our LJ Systems in Ref.~\onlinecite{fiocco2013oscillatory} and models of particle suspensions studied in \cite{keim2011generic}.

\subsection{Dynamical transition under oscillatory deformation:  the TM model}

\begin{figure}
			\includegraphics[width = 0.5\textwidth]{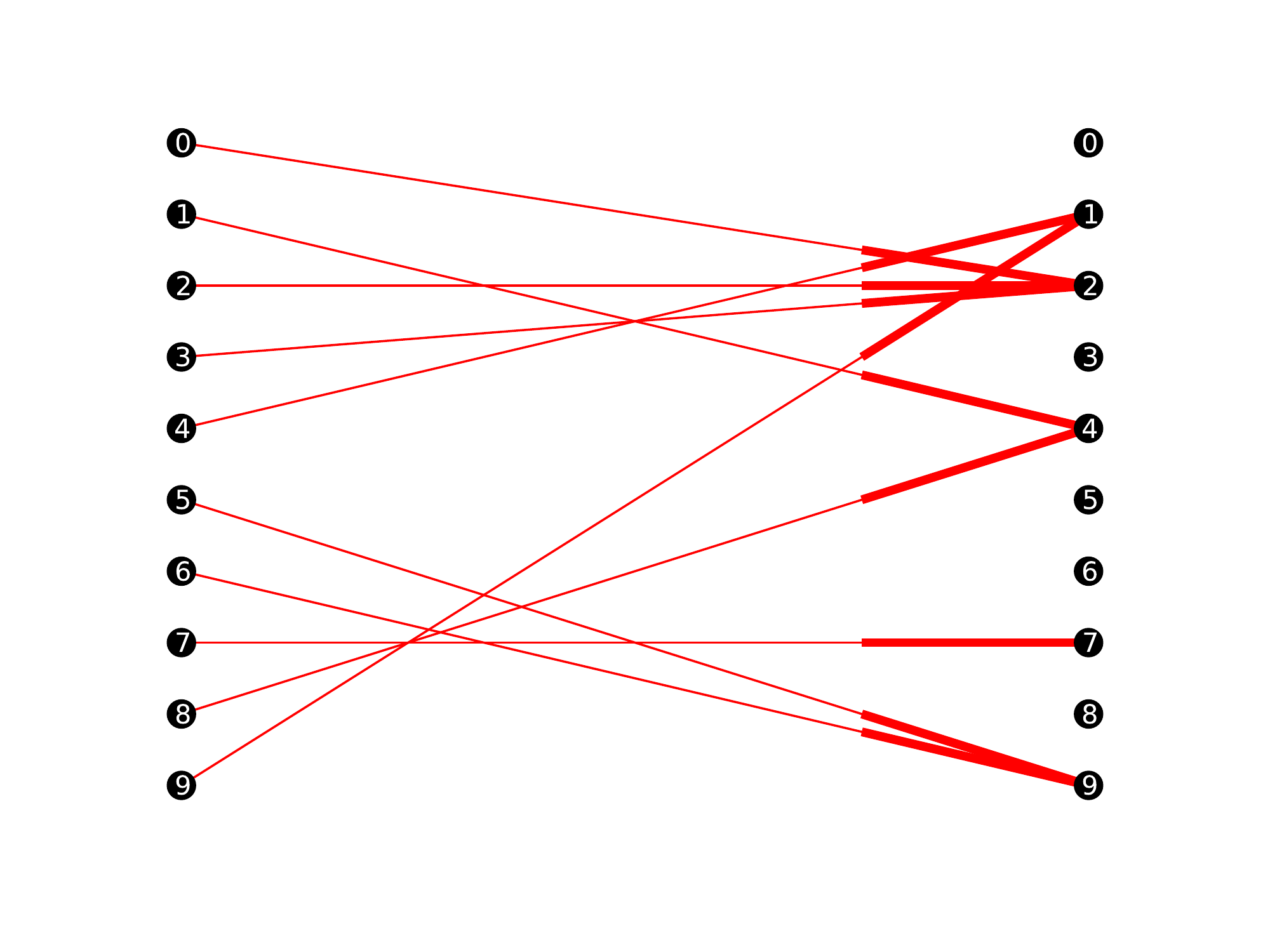}			
			\includegraphics[width = 0.5\textwidth]{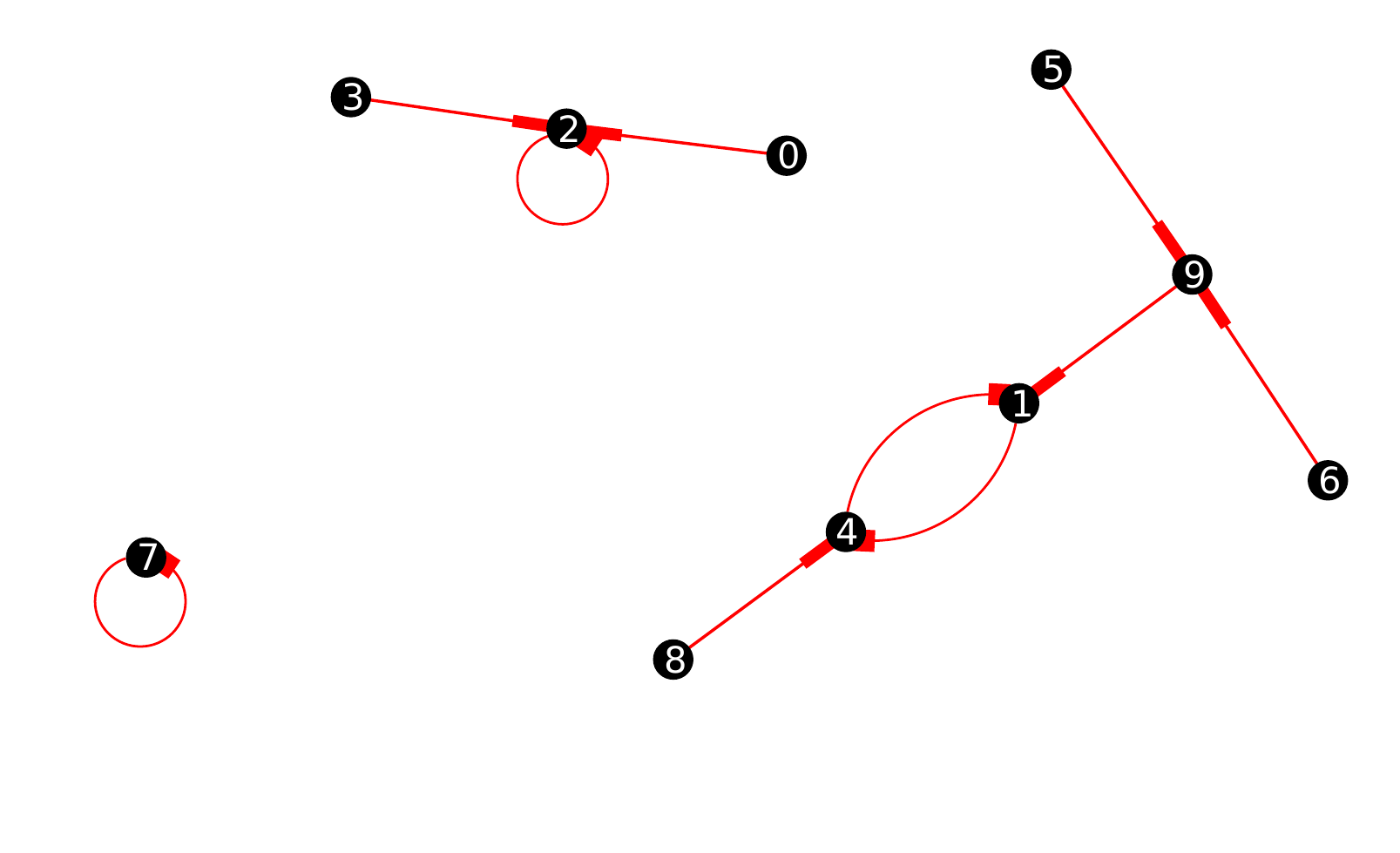}
	\caption{(Top) The output of the TM model is a map (which depends on $\gamma_{max}$) of the set of inherent structures onto itself. The same map can be interpreted as an adjacency matrix for a directed graph. (Bottom) The resulting graph is a collection of 1-trees \cite{bollobas1998modern}. Each of these 1-trees can contain a self-loop or not. If it does, then the vertex with the self-loop represents an absorbing state, and all the vertices in the 1-tree to which it belongs represent states mapping to that absorbing state via AQS dynamics. If the 1-tree does not contain self-loops, then its vertices are associated either to recurring states or states that map onto recurring states in the AQS dynamics. By counting the number of vertices in the two kinds of 1-trees (those with self-loops and those without), one can thus determine the fraction of states that are absorbing (and mapping to absorbing) or recurring (and mapping to recurring).\label{fig:DirectedGraphFromP}}
\end{figure}

Since the transition matrix model does not contain any real or configuration space distance information, a suitable measure of localization has to chosen for this model based on information about the fate of individual inherent structures under repeated operation of the transition matrix. 
The information contained in $P$ can be used to distinguish states that are absorbing or mapping to absorbing states from those that are recurring or mapping to recurring ones (see the definitions given in \autoref{sec:TMStateClassification}). This information, in turn, can be used to gather information about the dynamics under AQS deformation: if absorbing states dominate, systems are likely to be trapped into them, similarly to what is observed in the LJ and NK models below $\gamma_{c}$; if recurring states dominate, systems have the capability of exploring the configuration space before returning to the same state point, analogously to what happens in the other models above $\gamma_{c}$.
\\\\
The naive way to determine how inherent structures behave under oscillatory deformation is to apply $P$ repeatedly to each of them. After some applications of $P$, the structures will transform into states of the kind $\mathbf{R_{abs}}$ such that $P \mathbf{R_{abs}} = \mathbf{R_{abs}}$ (absorbing), or states of the kind $\mathbf{R_{rec}}$ such that $P^{L} \mathbf{R_{rec}} = \mathbf{R_{rec}}$ (recurring). This procedure is however too expensive, because one needs to calculate the trajectory of each and every of the $M$ structures by means of matrix multiplication. 
A way to overcome this is treating $P$ as an adjacency matrix, and constructing the directed graph\footnote{A useful reference for the terminology of graph theory used in this paragraph is \cite{bollobas1998modern}.} $G$ associated to it (see \autoref{fig:DirectedGraphFromP}). $G$ will be a directed graph whose outdegree is 1, as each structure maps onto one and only one configuration through $P$. In general, $G$ will possess several connected components\footnote{Using the terminology of graph theory, $G$ is a \emph{directed 1-forest} or a \emph{functional graph} and its connected components are called 1-trees \cite{bollobas1998modern}.}. Each of these either contains a self-loop or not. Connected components containing a self-loop are those that contain an absorbing state, which is a node that is connected to itself via the self-loop. All the other nodes are connected to it, and thus represent states mapping to the absorbing state. Connected components not containing self-loops must contain a loop, and their nodes thus represent recurring states or states mapping to recurring states. By examining the graphs one can count the number $R$ of recurring (and mapping to recurring) states simply by counting the number of nodes of the connected components of $G$ which do not possess self-loops. The number of absorbing (and mapping to absorbing) states will be given by $M-R$.

\begin{figure} 
\centering 
\includegraphics[width=0.5\textwidth]{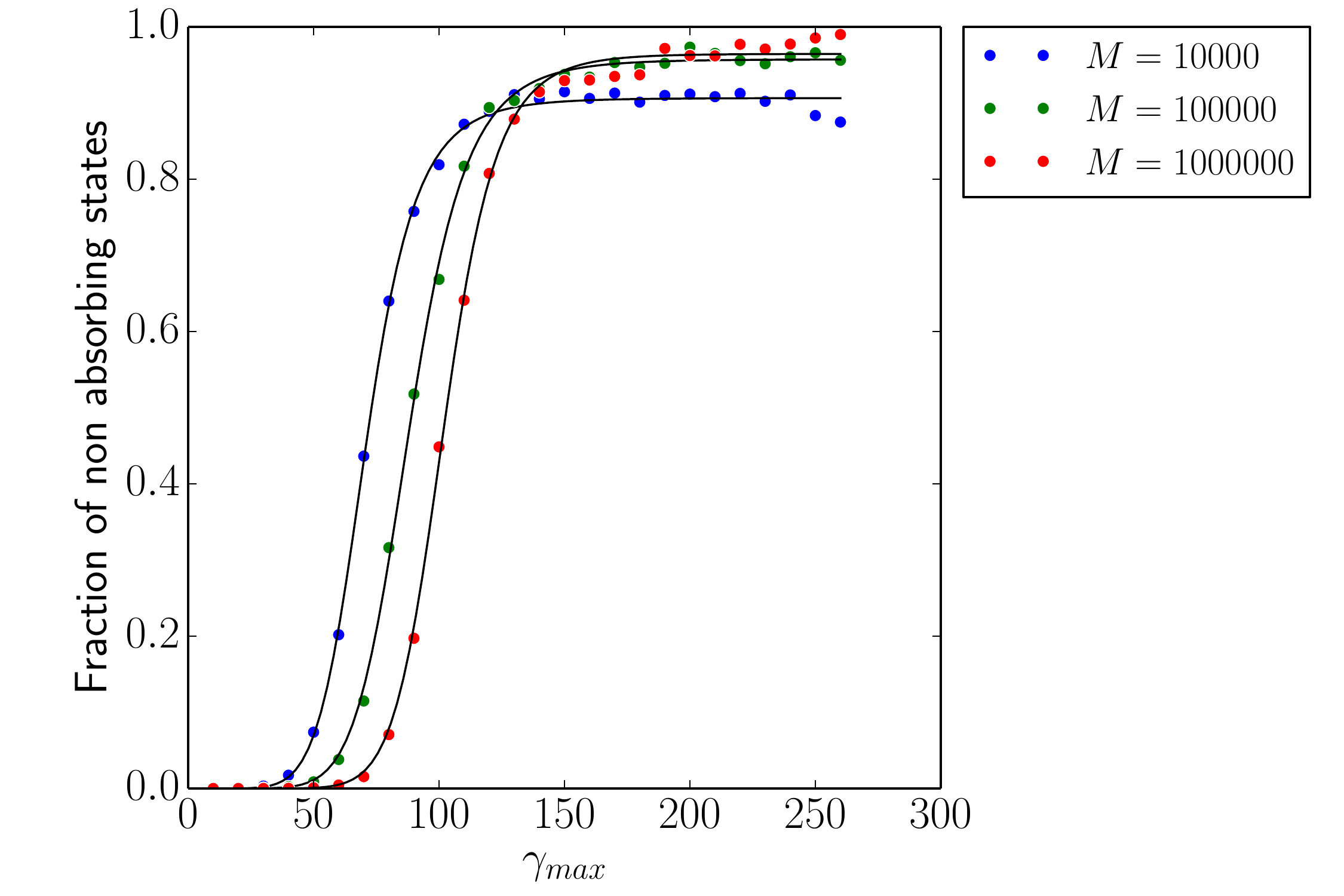} 
\caption{Overall fraction of recurring states and mapping to recurring states as a function of $\gamma_{max}$ for different values of $M$, obtained by analysis of the graphs associated to the transition matrices $P$ generated by the TM model. The number of recurring states increases strongly beyond some value $\gamma_{c}$ which increases as $M$ increases. The data can be described fairly well with the model in \autoref{eq:SaturatingPowerLaw} (black curves), with a sharpness (dictated by the parameter $a$) which increases for larger $M$. \label{fig:RecurrentStateFractionVsGammaMax}}
\end{figure}

We obtain $P$ with the procedure described in \autoref{append}, using Python and the support for sparse matrices within the library SciPy \cite{scipy}. We then extract $G$ and its connected components using NetworkX \cite{networkx}. The connected components associated to recurring states can be easily filtered because they don't contain self-loops. We do so for matrices $P$ with $M = 10^{4}$, $10^{5}$ and $10^{6}$, setting (tentatively) the probability for an inherent structure to be destabilized per unit strain to $\tau = 0.04$ and plot the average fraction of recurring (or mapping to recurring) states as a function of the $\gamma_{max}$ averaging on $\approx 800, 200, 50$ matrices respectively.
The result is shown in \autoref{fig:RecurrentStateFractionVsGammaMax}.  

The curves in \autoref{fig:RecurrentStateFractionVsGammaMax} can be modeled by the fitting function:
\begin{equation}
	f(x) = \frac{b}{1 + \left(\frac{\gamma_{c}}{\gamma_{max}} \right)^{a}}
	\label{eq:SaturatingPowerLaw}
\end{equation}

Data in \autoref{fig:RecurrentStateFractionVsGammaMax} and the form of \autoref{eq:SaturatingPowerLaw} show that the TM model shows a sharp increase in the number of states mapping to non-absorbing states as the ``oscillation amplitude'' is increased beyond some value $\gamma_{c}$, similarly to what has been observed in the case of LJ mixtures and of the NK model above. For this reason, one can reasonably believe that while crude, the TM model capture enough details to describe qualitatively the transition from a ``localized'' regime (where absorbing states prevail) to a ``diffusive'' one (where recurring states dominate) observed in particle models. Moreover, the transition is observed to be sharper for higher values of $M$, with the parameter $a$ in \autoref{eq:SaturatingPowerLaw} increasing with increasing $M$. Opposite to what is observed in LJ systems, however, the value of $\gamma_{c}$ is seen to increase with the system ``size'' $M$. It would be useful to address the question of whether a sharp transition arises in the thermodynamic limit rigorously \cite{During2009}.

	\section{Results: memory effects}
	\label{results-memory}
\begin{figure} 
\centering 
\includegraphics[width=0.5\textwidth]{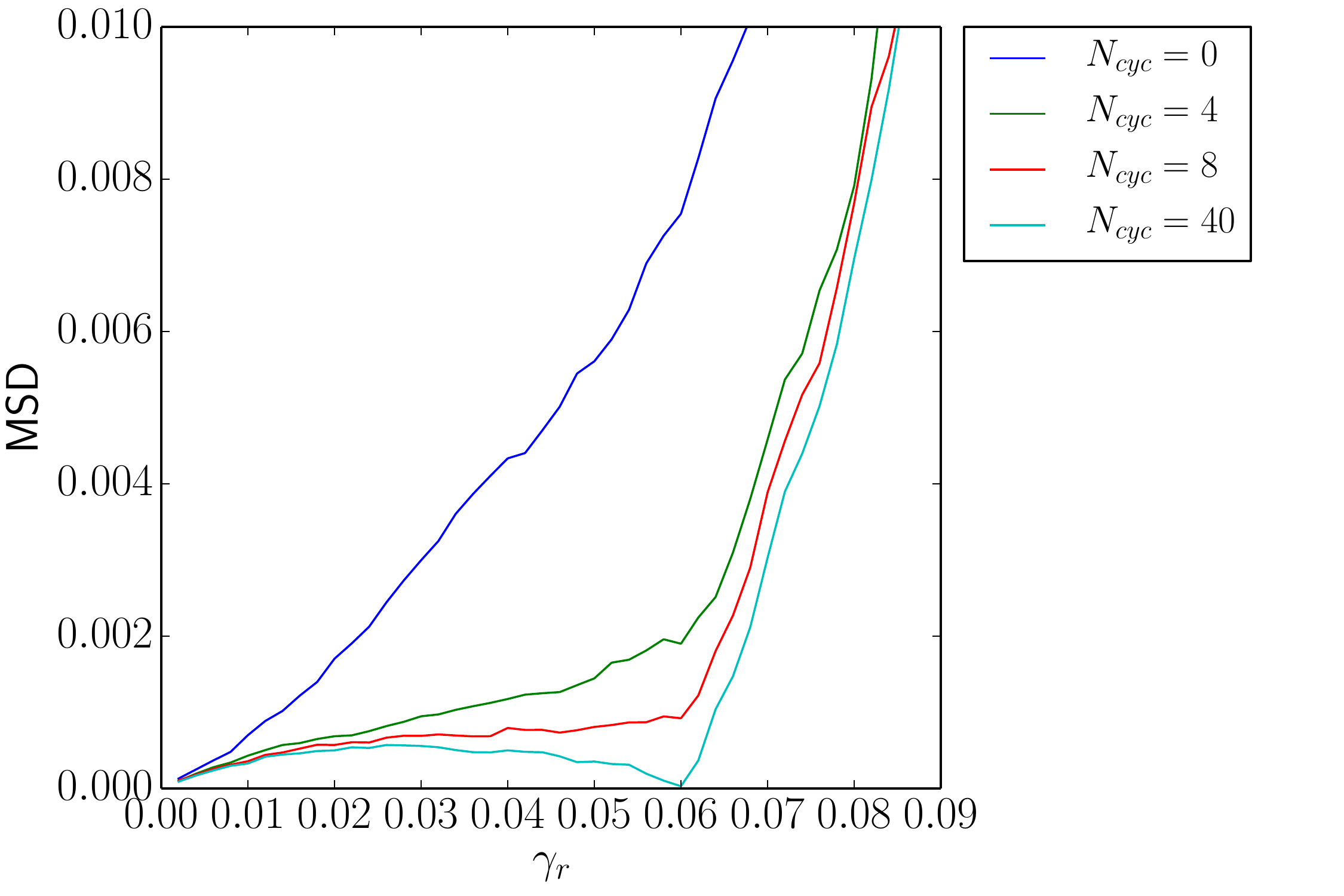} 
\includegraphics[width=0.5\textwidth]{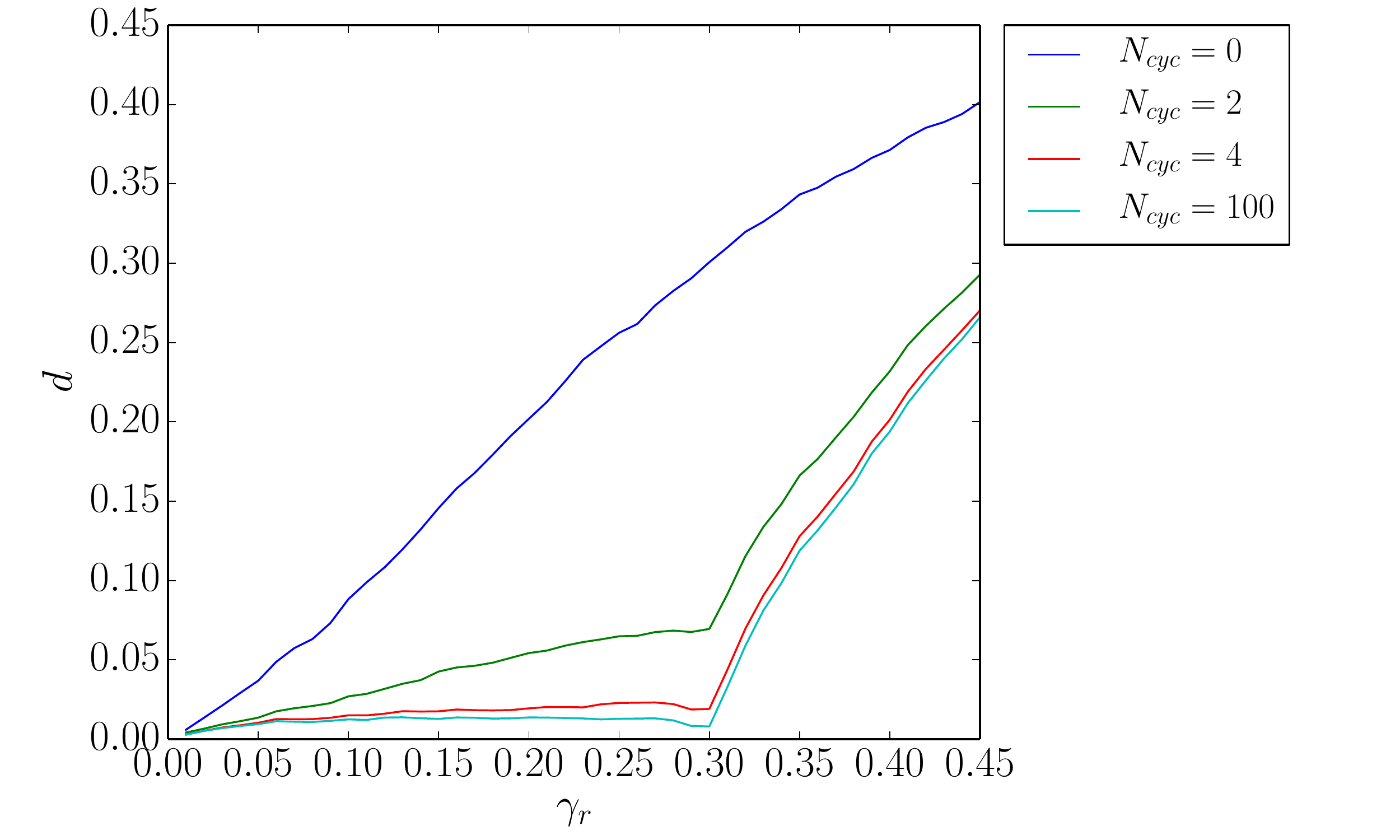} 
\includegraphics[width=0.5\textwidth]{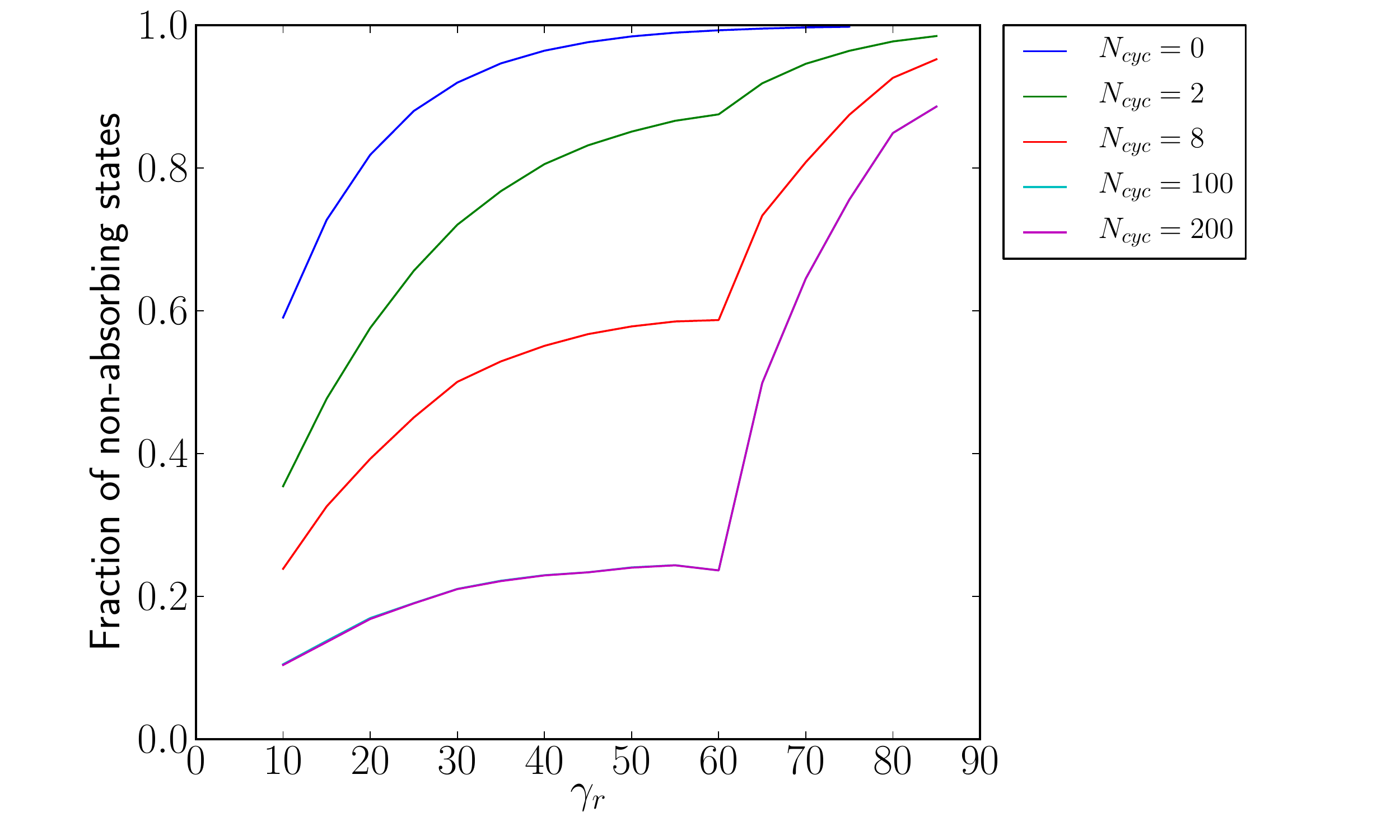} 
\caption{{\bf Single memory:} (Top) KA model as discussed in Ref. ~\onlinecite{fiocco2014encoding}. (Center) NK  model as discussed in Ref.~\onlinecite{fiocco2014encoding}. (bottom) Fraction of inherent states that are not invariant under the application of a $P_{\gamma_{r}}$, starting from a pool of states trained by a different number of applications of the matrix $P_{\gamma_{1}}$ with $\gamma_{1} = 60$, as a function of $\gamma_{r}$. Data are obtained within the TM model setting $M=10000$. It's clear how trainings of increasing length yield samples that show a memory of the training amplitude.  \label{fig:SingleMemoryTM}}
\end{figure}

In earlier work \cite{fiocco2014encoding} the memory effects in a model atomic glass former (LJ) and the NK model were studied, and shown to be similar, and we showed that both systems were capable of encoding persistent multiple memories. 
We thus focus here on the memory effects found in the TM model, but compare them with the behavior for the cases studied earlier. 
The procedure to probe memory effects in the TM model is fairly different with respect to that followed in the LJ and NK cases. 
First of all, we generate $P$ matrices for different values of $\gamma_{max}$, following the procedure described in \autoref{append}.
Once a $\gamma_{1}$ is chosen, we consider the configurations trained by $N_{cyc}$ oscillations (equivalent to $\gamma_{acc} = 4\gamma_{1} N_{cyc}$). 
These are configurations for which the corresponding rows  of the matrix $P_{\gamma_{1}}^{N_{cyc}}$ have at least one non-zero entry, where $P_{\gamma_{1}}$ is the matrix associated to the deformation up to $\gamma_{1}$. The  configurations that, after the $N_{cyc}$  of training, correspond to rows with all zero entries in the deformation matrix $P_{\gamma_{1}}^{N_{cyc}}$ cannot be reached by any further transformation and cannot be considered.  The remaining ones are those that can be still trasformed under the effect of the deformation.
We call to the set of such states $A_{N_{cyc}}$.
To probe the behavior of such states under a single reading cycle of amplitude $\gamma_{r}$, we check whether they are absorbing cycles for $\gamma_{r}$, i.e. we verify if the condition $P_{\gamma_{r}} \mathbf{R} = \mathbf{R}$ is satisfied, for each $\mathbf{R}$ state in $A_{N_{cyc}}$. The validity of this condition is easy to check, because it holds true if and only if the matrix element on the diagonal of $P_{\gamma_{r}}$ corresponding to such states is equal to one. In \autoref{fig:SingleMemoryTM} we plot the fraction of non-absorbing states (i.e. $\mathbf{R}$ states in $A_{N_{cyc}}$ that don't meet the condition $P_{\gamma_{r}} \mathbf{R} = \mathbf{R}$) for different values of $\gamma_{r}$, obtained by studying the states belonging to a space with $M= 10000$ structures with probability $\tau = 0.04$ to be destabilized and ``trained'' by a different number of cycles of amplitude $\gamma_{1} = 60$.

The results shown in \autoref{fig:SingleMemoryTM} indicate that the TM model displays single memory effects very similar to the LJ and NK models. 
Strictly speaking, the plot in \autoref{fig:SingleMemoryTM}(bottom) is not equivalent to those in \autoref{fig:SingleMemoryTM}(top) and \autoref{fig:SingleMemoryTM}(center): in the LJ and NK models a notion of distance exists between configurations, so that one is able to quantify the displacement experienced by the samples during the reading phase (using the MSD and the Hamming distance respectively); such information is not available for the TM model, which however offers very similar information: we can answer whether states are left unchanged or are modified by a reading cycle of amplitude $\gamma_{r}$\footnote{In this sense, the TM model gives a ``yes or no'' information about whether a sample is changed by a reading cycle; the LJ and NK models do more: they give a measure of \emph{how much} a sample is affected by a reading cycle.}.

\subsection{Multiple memories}

\begin{figure} 
\centering 
\includegraphics[width=0.5\textwidth]{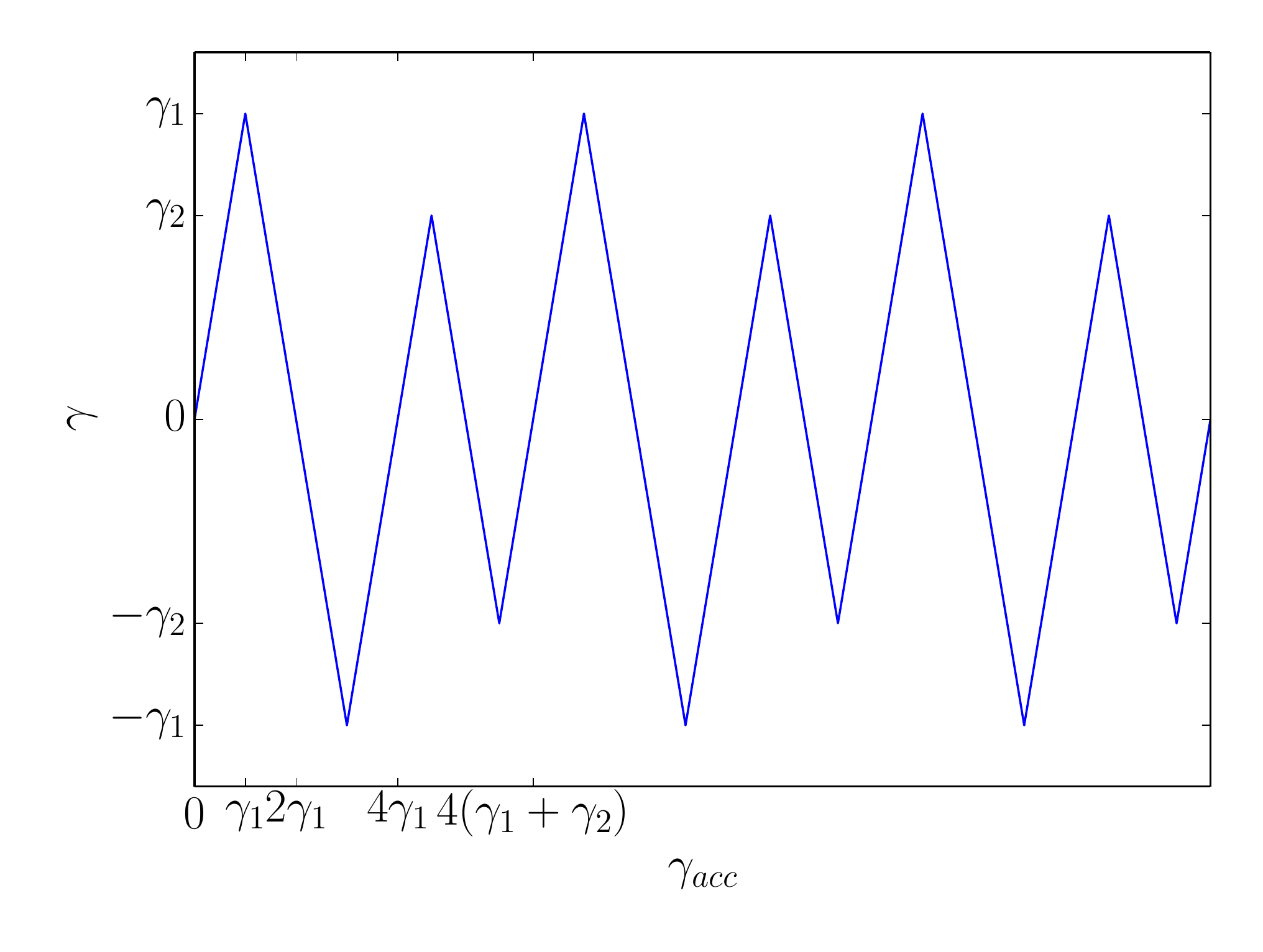} 
\caption{Strain profile applied to a sample to encode a double memory.\label{fig:DoubleTriangleWave}}
\end{figure}

\begin{figure} 
\centering 
\includegraphics[width=0.45\textwidth]{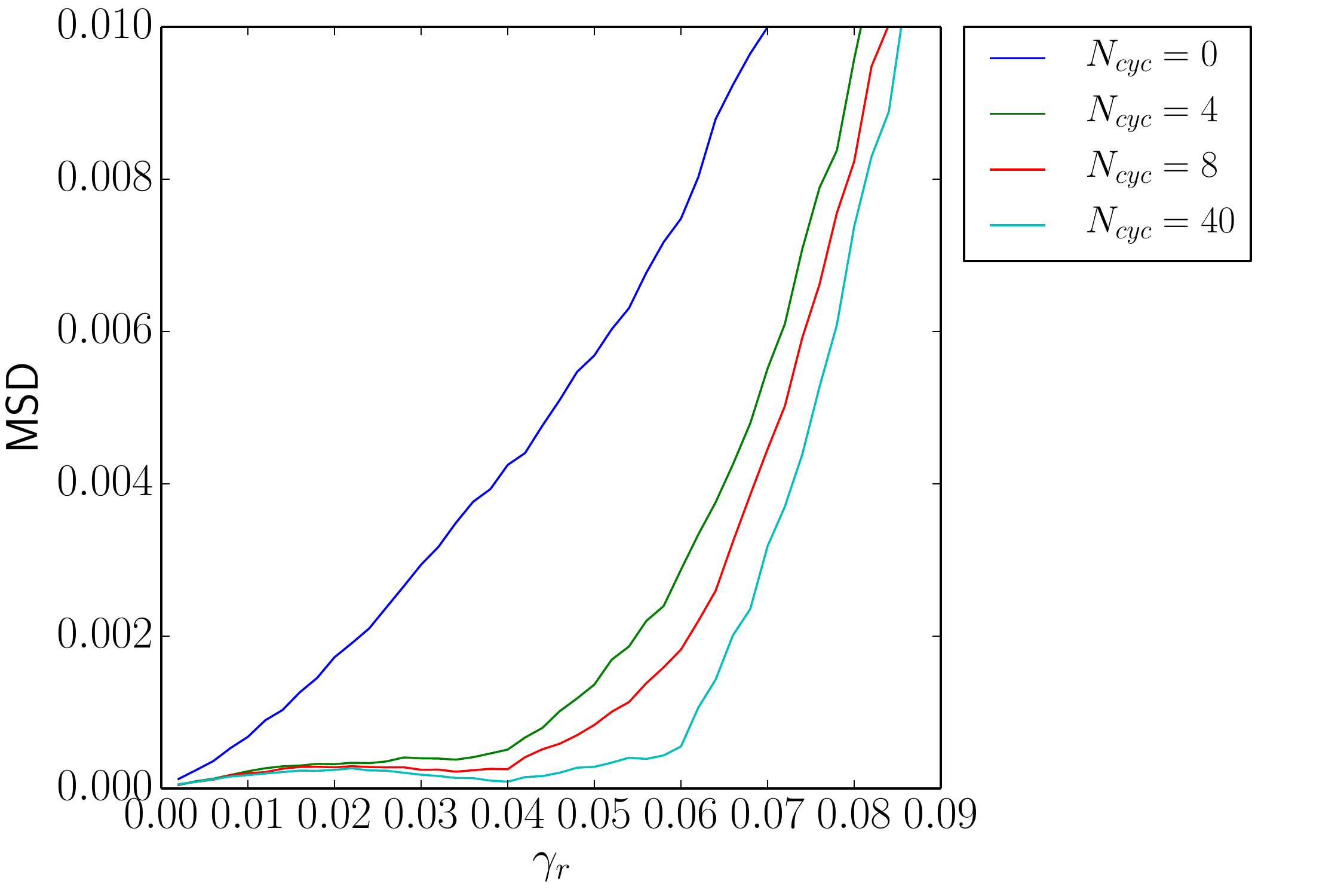} 
\includegraphics[width=0.45\textwidth]{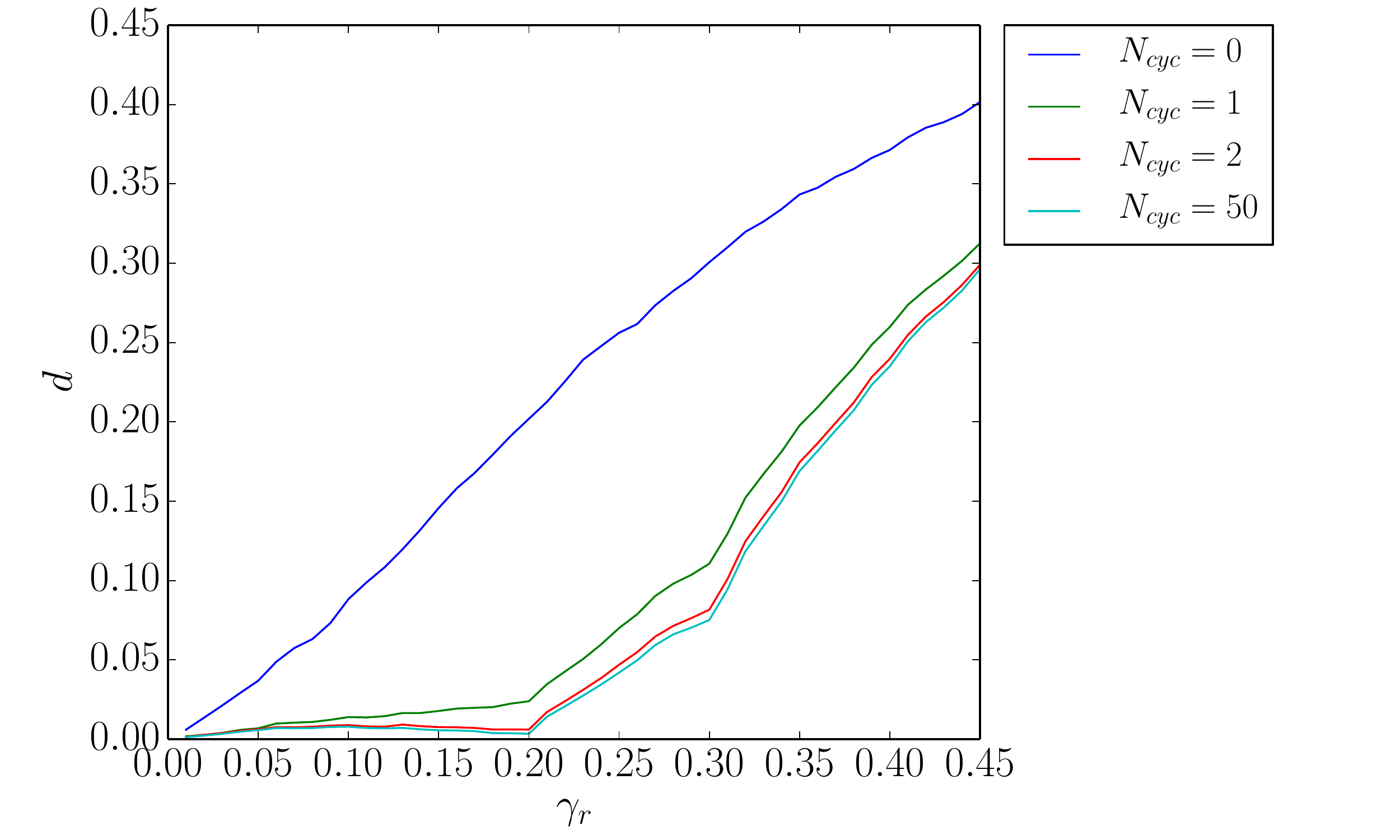} 
\includegraphics[width=0.45\textwidth]{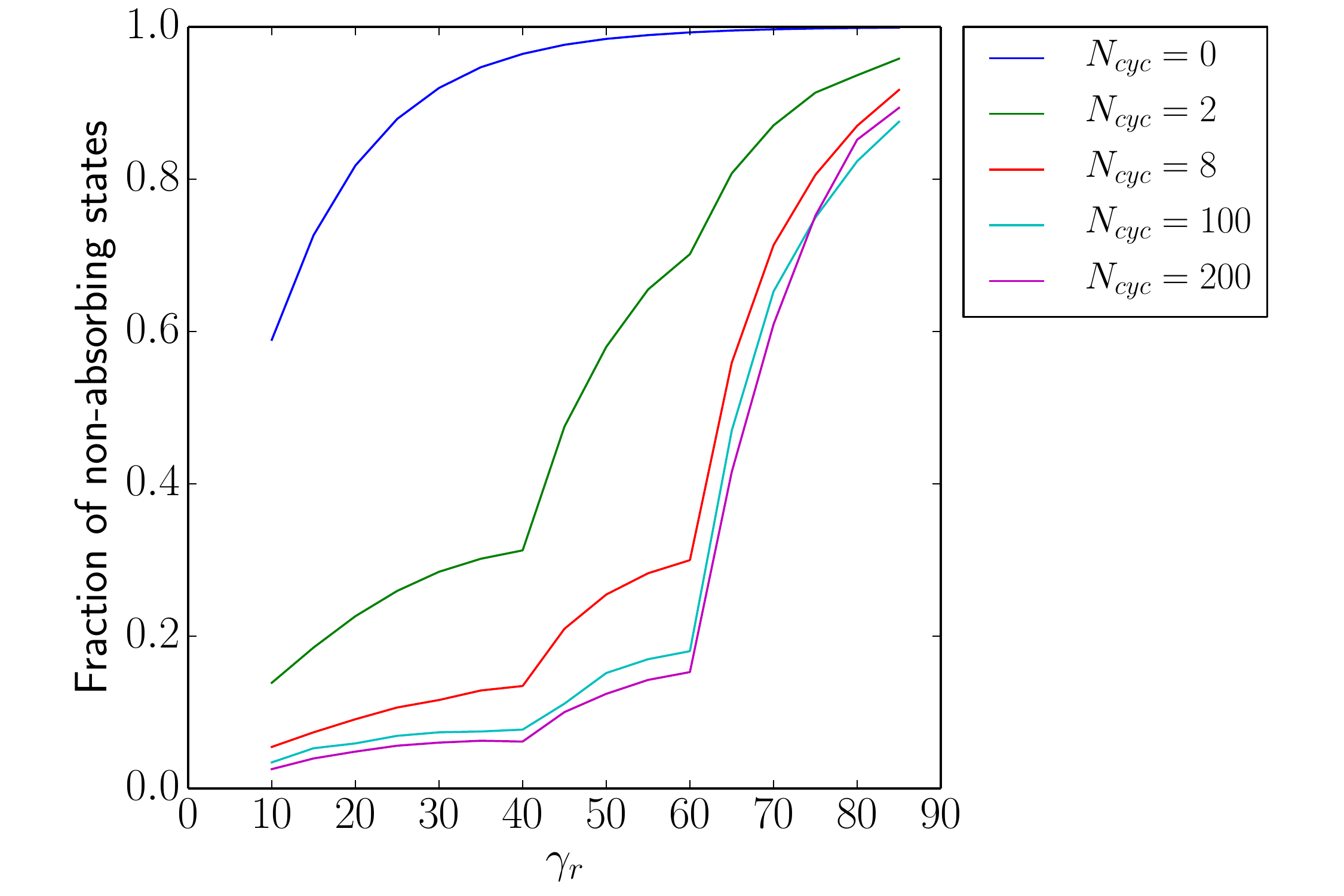} 
\caption{(Top) \textit{KA model:} Mean squared displacement between configurations before and after a full deformation cycle of amplitude $\gamma_{r}$, for a different number of training cycles alternating training amplitudes $\gamma_{1} = 0.06$ and $\gamma_{1} = 0.04$, as a function of $\gamma_{r}$. Data are relative to initially undeformed KA samples with $N=4000$ and whose effective temperature is $T=0.466$ (in reduced units see \cite{fiocco2014encoding} for further details). In the case of the longest trainings samples show a memory of both the training amplitudes. (Center){\it NK model:} Hamming distance between between configurations before and after a full deformation cycle of amplitude $\gamma_{r}$, for a different number of training cycles alternating training amplitudes $\gamma_{1} = 0.3$ and $\gamma_{1} = 0.2$, as a function of $\gamma_{r}$. Data are relative to initially undeformed NK samples with $N=20$ and whose effective temperature is $T=1.0$. In the case of the longest trainings samples show a memory of both the training amplitudes. (Bottom) \textit{TM model:} Fraction of inherent states that are not invariant under the application of a $P_{\gamma_{r}}$, starting from a pool of states trained by a different number of applications of the matrix $P_{\gamma_{1}}$ with $\gamma_{1} = 60$ and of $P_{\gamma_{2}}$ with $\gamma_{2} = 40$, as a function of $\gamma_{r}$. Data are obtained within the TM model setting $M=10000$. It's clear how trainings of increasing $N_{cyc}$ produce samples that show a memory of the training amplitudes. \label{fig:DoubleMemory}.}
\end{figure}

The training can be modified so that it consists of alternated repetition of cycles of amplitude $\gamma_{1}$ and $\gamma_{2}$, like in \autoref{fig:DoubleTriangleWave}. The rationale behind this protocol is to encode \emph{multiple} memories in our samples, and be able to read the values of the different training amplitudes in the reading phase.
The cycles in $\gamma$ have the form $0 \rightarrow \gamma_{1} \rightarrow 0 \rightarrow  -\gamma_{1} \rightarrow 0 \rightarrow \gamma_{2} \rightarrow 0 \rightarrow  -\gamma_{2} \rightarrow 0$. 
We repeat such cycle $N_{cyc}$ times, so that after the training the sample as been subjected to an accumulated strain $\gamma_{acc} = 4(\gamma_{1} + \gamma_{2})N_{cyc}$. This can be straightforwardly done in the LJ and NK cases, whereas (as described above in the case of single memory) a different scheme must be adopted with the TM model. We compare results for the LJ, NK and TM models.\\
For the LJ model, we choose $\gamma_{1} = 0.06$ and $\gamma_{2} = 0.04$ and train samples of the same size and initial effective temperature as those trained with a single amplitude by performing $N_{cyc}$ on them. We then take copies of the trained samples and subject them to a reading cycle of amplitude $\gamma_{r}$. As above we measure the MSD of the configurations as a function of $\gamma_{r}$. As it can be seen from \autoref{fig:DoubleMemory}(top), the MSD has two kinks in correspondence of $\gamma_{1}$ and $\gamma_{2}$, which are both visible for sufficiently high $N_{cyc}$. In addition, for a high number of $N_{cyc}$, the MSD curve converges to a curve showing clearly the trace of the two training amplitudes. By looking at the data, it is reasonable to assume that this will be true for an arbitrarily large number of $N_{cyc}$, so that the two memories will be \emph{persistent} for $\gamma_{acc} \rightarrow \infty$.

For the NK model, we choose $\gamma_{1} = 0.06$ and $\gamma_{2} = 0.04$ and train the samples for different $N_{cyc}$ with the same $N$, $K$, initial effective temperature and values of the couplings of those trained with a single amplitude. Again, we use the Hamming distance $d$ as a measure of the change in configurations in the reading phase and plot $d$ for different values of $\gamma_{r}$ in \autoref{fig:DoubleMemory}(center). As in the LJ case, the training amplitudes can be read by looking at kinks (discontinuities in the first derivative) in the plot, and again they appear persistent in the limit $\gamma_{acc} \rightarrow \infty$.

In the case of the TM model the ``trained'' configurations are those with the same index of the non-empty rows of the matrix $(P_{\gamma_{2}} P_{\gamma_{1}})^{N_{cyc}}$, where $P_{\gamma_{1}}$ and $P_{\gamma_{2}}$ are the matrices associated to the deformation up to $\gamma_{1}$ and $\gamma_{2}$. We refer to the set of such states as $B_{N_{cyc}}$. Exactly as in the reading of single memories, we compute the fraction of non-absorbing states (i.e. states in $B_{N_{cyc}}$ that don't meet the condition $P_{\gamma_{r}} \mathbf{R} = \mathbf{R}$) for different values of $\gamma_{r}$) as a function of $\gamma_{max}$.
From the analysis of \autoref{fig:DoubleMemory} (bottom), it is clear that a double memory can be encoded in an ensemble of structures in the TM model.
	
\section{Conclusions}
\label{conclude}
In this paper we have investigated the role of oscillatory deformation in two \textit{toy models} that have been introduced to mimic the behavior of a model of glass former that we studied before~\cite{fiocco2013oscillatory,fiocco2014encoding}. In particular we were interested in the presence of a dynamical transition from a localized to a diffusive state upon increasing the amplitude of the oscillations and in the encoding of memory in the system when a proper training protocol is followed.\\
The first model that we discussed is the NK model, a disordered spin model that has been introduced to investigate the deformation in glasses~\cite{isner2006generic}. The main property of the NK model is an energy landscape with tunable roughness. We have shown in our previous work that in this model we could induce a memory effect by oscillatory training cycles in a similar way to what can be achieved in LJ glass formers~\cite{fiocco2014encoding} and colloids~\cite{keim2011generic}.  Here we have discussed in more detail the nature of this energy landscape and we have shown that under oscillatory deformations, the NK model presents a dynamical transition at a critical amplitude of strain that is very similar to the one that is found in glass formers~\cite{fiocco2013oscillatory}. 
\\ 
We also introduced a second  model, the TM model. This model is based on a matrix approach that represents a further abstraction with respect to the NK model. The main idea is that, after a deformation cycle, when the sample returns to a state of zero deformation, the potential energy landscape is fixed. Consequently, one can map the act of deforming into the action of a transition matrix that changes the occupied minima of the landscape that are represented in a vector space. We have shown that, by imposing a few hypotheses, a reasonable transition matrix can be built. Despite its simplicity, we have shown that this model is capable of reproducing both the dynamical transition and the memory effects that we have found in the LJ and NK models~\cite{fiocco2013oscillatory,fiocco2014encoding}.\\
In conclusion, the NK and TM models are interesting for two main reasons. Firstly, they clarify what are the essential model features behind the dynamical transition observed in the LJ model. The NK model, for example, proves that a continuous configuration space and energy landscape is not necessarily required. The TM model goes even further and shows that the observed phenomena occur when the transition matrix is constructed following a simple set of rules. Having found that such a set of basic ingredients exist, one can imagine that analogous dynamic and memory effects could be observed in a wide class of systems, such as spin models or soft matter materials. Secondly, these simplified models could be the base of more fundamental approaches to treat the athermal deformation of physical system, in which the dynamics is deterministic. The TM model, for example, could lay down the basis for a new framework to model oscillatory deformation in materials. 
\\
The two  models that we have described here make some fundamental simplifications on the properties of the PEL. The TM model assumes that the states visited during deformation are completely uncorrelated (as there is no notion of space or distance in the TM model), while the NK introduces a PEL of controlled roughness but with no structural information, as it is generated from a series of random couplings. This is fundamentally different from the full KA model that presents a PEL that is the result of a two-body interaction potential between its constituents. A present limitation of the TM model, however, is that it doesn't incorporate any organization between the different inherent states. This is clearly not the case for the LJ systems, whose PEL has a complex structure and spatially correlated particle motion between adjacent inherent structures is expected under deformation. In the future, it would be interesting to see if some degree of correlation can be included in this model. Another interesting question for future work is the identification of precise conditions for persistent memory as seen in the NK and TM models, as opposed to transient memory seen in other model systems \cite{keim2011generic}. 

\begin{acknowledgments}
We thank  S. Franz,  J. Kurchan, S. R. Nagel, E. Vincent, and T. Witten for illuminating discussions and F. Varrato for the careful reading of the manuscript. We acknowledge support from the Indo-Swiss Joint Research Programme (ISJRP). D. F. and G. F. acknowledge financial support from Swiss National Science Foundation (SNSF) Grants No. PP0022\_119006 and No. PP00P2\_140822/1.
\end{acknowledgments}
\newpage

\appendix
\section{Details of the construction of $P$}
\label{append}

Here we describe how the assumptions listed in \autoref{sec:TMAssumptions} can be used to construct $P$.
First of all, one can see that
\begin{equation}
    P = P_{-} P_{+} 
\end{equation}
so that the construction of $P$ is reduced to that $P_{+}$ and $P_{-}$. 
Each of these (say $P_{+}$), can in turn be viewed as the composition of 
matrices 
\begin{widetext}
\begin{equation}
P_{+} =
P_{\leftarrow}^{0}
\ldots
P_{\leftarrow}^{\gamma_{max} - d\gamma}
P_{\rightarrow}^{\gamma_{max}}
P_{\rightarrow}^{\gamma_{max} - d\gamma}
\ldots
P_{\rightarrow}^{2d\gamma}
P_{\rightarrow}^{d\gamma}
\end{equation}
\end{widetext}
where $P_{\rightarrow}^{\gamma^{*}}$ is the matrix describing how AQS dynamics maps the inherent structures of the landscape relative to $\gamma = \gamma^{*} - d\gamma$ into the set of the structures of the landscape associated to $\gamma = \gamma^{*}$. The arrows in the subscript indicate whether $P^{\gamma^{*}}$ is associated to an increase of $\gamma$ ($\rightarrow$) or a decrease ($\leftarrow$) in strain. For instance, $P_{\leftarrow}^{\gamma^{*}}$ describes how AQS maps the structures of the $\gamma = \gamma^{*} + d\gamma$ landscape onto those associated to $\gamma = \gamma^{*}$. 
As the landscape is assumed to have $M$ inherent structures no matter the value of $\gamma$, each of these $P_{\rightarrow}^{\gamma^{*}}$, $P_{\leftarrow}^{\gamma^{*}}$ is a square matrix.
To construct each of the $P_{\rightarrow}^{\gamma^{*}}$ one uses the assumption that the probability per unit strain to destabilize a given inherent structure is equal to $\tau$. So, when the strain is incremented by $d\gamma$, a system in a given inherent structure has probability $1 - \tau d\gamma$ to be in a structure $i$ that is not destabilized, and thus it maps to the same structure $\mathbf{R_i}$ in the deformed landscape through $P_{\rightarrow}^{\gamma^{*}}$. In that case the matrix element $P_{\rightarrow, ii}^{\gamma^{*}} = 1$. The system has also a probability $\tau d\gamma$ to be in a structure $\mathbf{R_j}$ that is destabilized by the strain increment, so that it falls onto some randomly chosen inherent structure $\mathbf{R_k}$ of the deformed landscape. in that case the matrix element $P_{\rightarrow, kj}^{\gamma^{*}} = 1$. Incidentally, for each configuration $\mathbf{R_l}$ that is destabilized at $\gamma^{*}$ as strain is increased, another configuration $\mathbf{R_m}$ is correspondingly created at that strain value. This means that $\mathbf{R_m}$ will be destroyed at $\gamma^{*}$ when the strain will be decreased (due to the symmetry of the landscapes). This is a constraint on the form of the matrix $P_{\leftarrow}^{\gamma^{*}}$: $P_{\leftarrow}^{\gamma^{*}}$ must be constructed by taking into account that the structures that are destroyed at $\gamma^{*}$ when incrementing $\gamma$ are exactly those created at $\gamma^{*}$ when decrementing $\gamma$.\\
The procedure outlined above is used to create a matrix $P_{+}$ associated to some $\gamma_{max}$. The $P_{+}'$ corresponding to another $\gamma_{max}'$ is simply given by 
\begin{widetext}
\begin{equation}
P_{+}' =\begin{cases} P_{\leftarrow}^{0}
\ldots
P_{\leftarrow}^{\gamma_{max}' - d\gamma}
P_{\rightarrow}^{\gamma_{max}'}
P_{\rightarrow}^{\gamma_{max}' - d\gamma}
\ldots
P_{\rightarrow}^{2d\gamma}
P_{\rightarrow}^{d\gamma}  &\mbox{if } \gamma_{max}' < \gamma_{max}  \\ \\
P_{\leftarrow}^{0}
\ldots
P_{\leftarrow}^{\gamma_{max} - d\gamma}
P_{\leftarrow}^{\gamma_{max}}
P_{\leftarrow}^{\gamma_{max} + d\gamma}
\ldots
P_{\leftarrow}^{\gamma_{max}' - d\gamma}
P_{\rightarrow}^{\gamma_{max}'} 
P_{\rightarrow}^{\gamma_{max}' - d\gamma} \notag\\
\ldots 
P_{\rightarrow}^{\gamma_{max} + d\gamma}
P_{\rightarrow}^{\gamma_{max}}
P_{\rightarrow}^{\gamma_{max} - d\gamma}
\ldots
P_{\rightarrow}^{2d\gamma}
P_{\rightarrow}^{d\gamma}  &\mbox{if } \gamma_{max}' > \gamma_{max}  \end{cases} 
\end{equation}
\end{widetext}

Using this procedure its thus possible to create a sequence of matrices relative to multiple values of $\gamma_{max}$. A similar procedure can be followed for $P_{-}$).

\bibliographystyle{apsrev4-1}

\end{document}